\documentclass[%
 reprint,
superscriptaddress,
 amsmath,amssymb,
 aps,
]{revtex4-2}

\usepackage{graphicx}
\usepackage{dcolumn}
\usepackage{bm}
\usepackage{capt-of}

\begin{document}

\preprint{APS/123-QED}

\title{The evolution of magnetism in a thin film pyrochlore ferromagnetic insulator}

\author{Margaret A. Anderson$^*$}
\affiliation{Department of Physics, Harvard University, Cambridge, MA, USA}

\author{Megan E. Goh$^*$}
\affiliation{School of Engineering and Applied Sciences, Harvard University, Cambridge, MA, USA}

\author{Yang Zhang}
\affiliation{The Rowland Institute at Harvard, Cambridge, MA, USA}

\author{Kyeong-Yoon Baek}
\affiliation{Department of Physics, Harvard University, Cambridge, MA, USA}

\author{Michael Schulze}
\affiliation{Leibniz Institute for Crystal Growth, Max-Born-Str. 2, D-12489 Berlin, Germany}

\author{Mario Brützam}
\affiliation{Leibniz Institute for Crystal Growth, Max-Born-Str. 2, D-12489 Berlin, Germany}

\author{Christoph Liebald}
\affiliation{EOT GmbH – Coherent, Struthstr. 2, D – 55743 Idar-Oberstein, Germany}

\author{Chris Lygouras}
\affiliation{Institute for Quantum Matter and Department of Physics and Astronomy, Johns Hopkins University, Baltimore, MD, USA}

\author{Dan Ferenc Segedin}
\affiliation{Department of Physics, Harvard University, Cambridge, MA, USA}

\author{Aaron M. Day}
\affiliation{School of Engineering and Applied Sciences, Harvard University, Cambridge, MA, USA}

\author{Zubia Hasan}
\affiliation{Department of Physics, Harvard University, Cambridge, MA, USA}

\author{Donald Walko}
\affiliation{X-ray Science Division, Advanced Photon Source, Argonne National Laboratory, Lemont, IL, USA}

\author{Hua Zhou}
\affiliation{X-ray Science Division, Advanced Photon Source, Argonne National Laboratory, Lemont, IL, USA}

\author{Peter Bencok}
\affiliation{Diamond Light Source, Didcot, Oxfordshire,
OX11 0DE, United Kingdom}

\author{Alpha T. N'Diaye}
\affiliation{Advanced Light Source, Lawrence Berkeley
National Laboratory, Berkeley, CA, USA}

\author{Charles M. Brooks}
\affiliation{Department of Physics, Harvard University, Cambridge, MA, USA}

\author{Ismail El Baggari}
\affiliation{The Rowland Institute at Harvard, Cambridge, MA, USA}

\author{John T. Heron}
\affiliation{Department of Materials Science and Engineering, University of Michigan, Ann Arbor, MI, USA}

\author{Sayed Koopayeh}
\affiliation{Institute for Quantum Matter and Department of Physics and Astronomy, Johns Hopkins University, Baltimore, MD, USA
}%
\affiliation{Department of Materials Science and Engineering, Johns Hopkins University, Baltimore, MD, USA
}%
\affiliation{Ralph O’Connor Sustainable Energy Institute, Johns Hopkins University, Baltimore, MD, USA
}

\author{Daniel Rytz}
\affiliation{EOT GmbH – Coherent, Struthstr. 2, D – 55743 Idar-Oberstein, Germany}

\author{Christo Guguschev}
\affiliation{Leibniz Institute for Crystal Growth, Max-Born-Str. 2, D-12489 Berlin, Germany}

\author{Julia A. Mundy}
\affiliation{Department of Physics, Harvard University, Cambridge, MA, USA}
\affiliation{School of Engineering and Applied Sciences, Harvard University, Cambridge, MA, USA}


\begin{abstract}
The pyrochlore vanadates 
are compelling candidates for next-generation dissipationless devices. Lu$_2$V$_2$O$_7$ and Y$_2$V$_2$O$_7$ are ferromagnetic insulators (T$_\text{c} \sim$ 70 K) that are believed to exhibit the magnon Hall effect and are expected to host topological magnons. Their completely dissipationless magnon edge states could be harnessed to realize low-power information transport in spintronic or magnonic devices. As a crucial step in the realization of devices, we synthesize the first thin films of pyrochlore Y$_2$V$_2$O$_7$ on isostructural Y$_2$Ti$_2$O$_7$ substrates and explore the evolution of their magnetic properties down to the ultrathin limit. All films are insulating ferromagnets with transition temperatures of up to the bulk value (T$_\text{c} \sim$ 68 K) that decrease with thickness according to finite-size effects. Our films also exhibit a change in anisotropy from in-plane to out-of-plane easy axis coincident with the development of partial strain relaxation and nonzero magnetic hysteresis in an applied field. This evolution demonstrates the impact of strain on magnetic anisotropy and paves the way to tunable magnon topology. 
\end{abstract}
\maketitle
\def\thefootnote{*}\footnotetext{These authors contributed equally to this work}\def\thefootnote{\arabic{footnote}}


\section{Introduction}\label{sec:intro}

The ferromagnetic insulating pyrochlore vanadates ($A_2$V$_2$O$_7$, $A$ = Y, Lu) are promising platforms for low-power spintronic \cite{shamoto2002substitution, pereiro2014topological, de2025observation} or magnonic \cite{onose2010observation, zhang2013topological, mook2015magnon, su2017magnonic, de2025observation} devices. In addition to interest as ferromagnetic insulators (T$_\text{c} \sim$ 70 K) \cite{shamoto2001light, ullah2017structural, xiang2011single, shamoto2002substitution,haghighirad2008powder,de2025observation}, the vanadate pyrochlores are predicted to host topological magnons with fully dissipationless magnon transport in edge states \cite{zhang2013topological, mook2014magnon, pereiro2014topological, mook2015magnon, su2017magnonic}. This nontrivial topology of magnons leads Lu$_2$V$_2$O$_7$ to exhibit the magnon hall effect \cite{onose2010observation, katsura2010theory, ideue2012effect, mook2014magnon, de2025observation} and suggests that the pyrochlore vanadates can transport information in the form of edge-confined, topologically-protected spin waves in low-power magnon-based devices. 

\begin{figure*}
    \centering
    \includegraphics[width=0.8\textwidth]{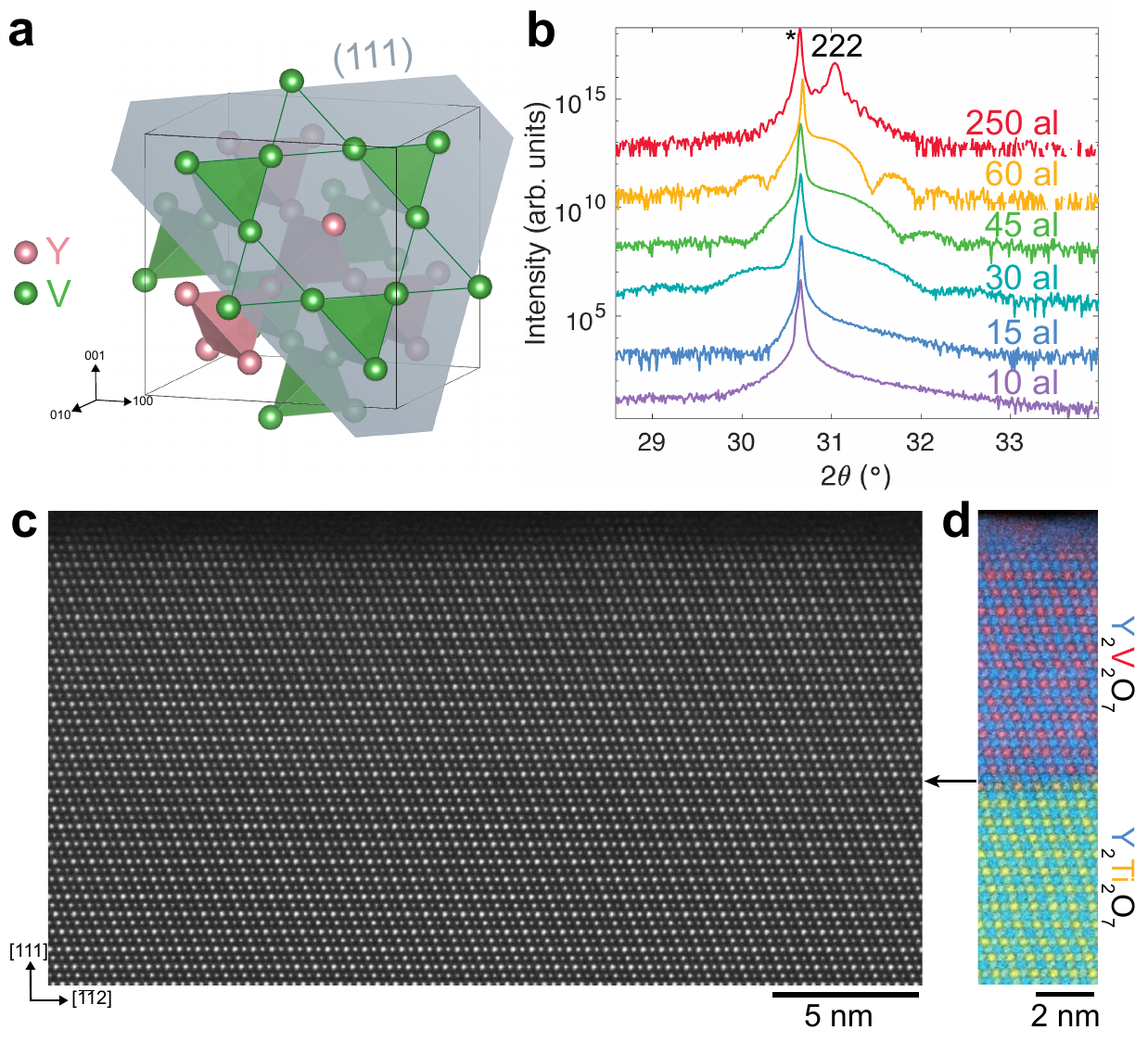}
    \caption{Pyrochlore structure and characterization. \textbf{a} The cation structure of Y$_2$V$_2$O$_7$ showing the separate interpenetrating pyrochlore sublattices of corner sharing tetrahedra and an isolated 111 vanadium Kagome plane \textbf{b} X-ray diffraction about the 222 substrate (Y$_2$Ti$_2$O$_7$, denoted by asterisk) and film (Y$_2$V$_2$O$_7$) for the thickness series (al = atomic layers) showing thickness fringes indicative of smooth, high-quality films \textbf{c} HAADF-STEM micrograph of the 30 atomic layer Y$_2$V$_2$O$_7$ film along $\langle1\overline{1}0\rangle$ showing nearly indistinguishable interface between substrate and film \textbf{d} An EELS micrograph of the same film as (c) at approximately the same scale showing a clear interface between film (top) and substrate (bottom)}
    \label{StructandChar}
\end{figure*}

The vanadates belong to a family of isostructural materials, the pyrochlore oxides, with generic formula $A_2B_2$O$_6$O' where both the $A$ and $B$ cations form separate but interpenetrating sublattices of corner sharing tetrahedra (Figure \ref{StructandChar}a) often called the pyrochlore sublattice \cite{gardner2010magnetic}. This sublattice, the 3D analog of the Kagome lattice, is geometrically frustrated.
In $A_2$V$_2$O$_7$, V$^{4+}$ (3$d^1$, s = $\frac{1}{2}$) sits in a nearly octahedral oxygen coordination environment with a slight trigonal distortion that splits the lower energy T$_\text{2g}$ states of the octahedral environment into a lowest lying A$_\text{1g}$ state and higher energy doubly-degenerate E$'_\text{g}$ states (Figure \ref{suppTrigCF}) \cite{biswas2013crystal, shamoto2002substitution, ichikawa2005orbital, xiang2011single}. In this configuration, the V$^{4+}$ ions host isotropic Heisenberg exchange and weak but nontrivial Dzyaloshinskii-Moriya (DM) superexchange allowed by Moriya's rules and enabled by spin-orbit coupling \cite{onose2010observation, xiang2011single, ideue2012effect, mena2014spin, riedl2016ab}. From this relatively simple model on the pyrochlore lattice, myriad unexpected properties emerge. Y$_2$V$_2$O$_7$ and isostructural Lu$_2$V$_2$O$_7$ are ferromagnetic insulators contrary to the expectation that the unpaired spins leading to a net magnetic moment should produce a metallic state \cite{shamoto2001light, ullah2017structural, xiang2011single, shamoto2002substitution}. The collinear ferromagnetism is thought to be stabilized by orbital order leading to nontrivial higher-order electron hopping \cite{miyahara2007orbital} or via thermal order-by-disorder that selects a subset of spin configurations from an O(3) manifold of degenerate states \cite{hickey2025order}. Although the DM interaction cancels out across nearest-neighbor bonds, preserving the collinear spin arrangement \cite{hickey2025order, onose2010observation, su2017magnonic}, the fictitious magnetic flux caused by ring exchange about inequivalent loops in the pyrochlore structure enables topological magnon edge states and the magnon Hall effect \cite{katsura2010theory, ideue2012effect, mook2014magnon}. Lu$_2$V$_2$O$_7$ powder exhibits a light-induced metal-to-insulator transition \cite{shamoto2001light} while single crystals show colossal magnetoresistance near the ferromagnetic transition temperature. Theoretical studies suggest device applications of pyrochlore vanadates ranging from a field-induced spin transistor \cite{shamoto2002substitution} to a topological magnon waveguide \cite{mook2015magnon}.

Early efforts to synthesize single crystals of the Y$_2$V$_2$O$_7$ yielded micron-scale crystallites embedded in a matrix of unreacted starting material \cite{haghighirad2008powder, haghighirad2008crystal}. Higher-quality floating zone single crystals of Lu$_2$V$_2$O$_7$ enabled the observation of orbital ordering \cite{ichikawa2005orbital}, ferromagnetic-cluster-mediated colossal magnetoresistance \cite{zhou2008magnetic}, and the first observation of the magnon Hall effect \cite{onose2010observation}. Recently, macroscopic single crystals of Y$_2$V$_2$O$_7$ have been reported as well \cite{de2025observation}. While single crystals are ideal for thermal transport and neutron scattering measurements, the thin film geometry is key to realizing practical devices. In addition to enabling device fabrication and integration, thin film synthesis offers a variety of pathways to modify and improve the functionality of materials. 
Film properties can be tuned with epitaxial strain, modulated doping, and interfacial effects \cite{schlom2008thin,nordlander2022epitaxy}. In thin films of Y$_2$V$_2$O$_7$, strain and dimensional confinement could modify the strength of magnetic exchange and tune the ferromagnetic ordering \cite{pereiro2014topological} or magnetic anisotropy, thereby impacting the topological magnon states \cite{su2017magnonic, jyothis2024magnon, mook2014magnon, mook2015magnon, chakhalian2020strongly}. 

Here, we synthesize the first thin films of Y$_2$V$_2$O$_7$ and characterize their magnetic properties as a function of thickness, strain, and dimensional confinement. We further take advantage of the Kagome-triangular plane layering in the [111] direction (Figure \ref{StructandChar}a) of the pyrochlore structure and the atomic precision of molecular beam epitaxy to isolate sub-unit-cell thicknesses and explore the unusual ferromagnetism of Y$_2$V$_2$O$_7$ in the ultrathin limit. We find that our high-quality thin films of pyrochlore Y$_2$V$_2$O$_7$ exhibit the ferromagnetic insulating state found in bulk crystals down to sub-unit-cell thicknesses. However, the Curie temperature of the films decreases in thinner films following finite-size effects. The thin films exhibit hysteresis under an applied field with a coercivity that increases with thickness. Finally, the Y$_2$V$_2$O$_7$ shows a change in anisotropy coincident with the onset of partial relaxation in thicker films and counter to fundamental expectations based on shape and surface anisotropy. Understanding the interplay of dimensionality, strain, and magnetism in these films is critical to the fabrication and realization of magnonic devices and sheds light on the intersection of topology and frustration in thin films writ large.

\section{Results}
\subsection{Synthesis and Structural Characterization}

Using reactive-oxide molecular beam epitaxy (MBE), we realize the first thin films of pyrochlore Y$_2$V$_2$O$_7$ on the isostructural substrate Y$_2$Ti$_2$O$_7$(111). YVO$_x$ on yttria-stabilized zirconia (YSZ) substrates forms a phase with pyrochlore-like reflection high energy electron diffraction (RHEED) and x-ray diffraction (XRD) patterns that indicate the expected doubling of the out-of-plane lattice parameter of the film compared to the substrate (Figure \ref{suppCharYVO_YSZ}). However, magnetic susceptibility measurements yield T$_\text{c}$ $\approx$ 140 K; twice as high as expected from bulk Y$_2$V$_2$O$_7$ crystals (Figure \ref{suppSquidYVO_YSZ}). Subsequently, high-angle annular dark field scanning transmission electron microscopy (HAADF-STEM) imaging reveals a non-pyrochlore phase with alternating 111 planes, but no discernible in-plane order (Figure \ref{suppSTEMYVO_YSZ}). Based on the film's fast Fourier transform (FFT) and comparison with the diffraction patterns of defect fluorite and pyrochlore crystals, we identify the phase as an anisotropic defect fluorite with pyrochlore-like ordering in the 111 direction. While this phase may be of interest as a highly insulating high-T$_\text{c}$ ferromagnetic insulator, it lacks the 111 vanadium kagome planes that lead to nontrivial magnon topology in Y$_2$V$_2$O$_7$. Therefore films on YSZ are not promising platforms for low dissipation magnonic devices.

Moving away from the more common YSZ substrate, we successfully stabilize pyrochlore Y$_2$V$_2$O$_7$ films on noncommercial Y$_2$Ti$_2$O$_7$ pyrochlore substrates (Figure \ref{suppYTOsub}). In situ RHEED indicates smooth, highly crystalline films (Figure \ref{suppRHEED}). Atomic force microscopy confirms that the film surfaces are atomically smooth with RMS roughness on the order of 150 pm (Figure \ref{suppAFMstoich}). As on YSZ, XRD exhibits the expected pyrochlore peaks with thickness fringes that are a further testament to film quality (Figure \ref{StructandChar}b and Figure \ref{suppXRDfull}). HAADF-STEM imaging shows an exceptionally high-quality film, nearly indistinguishable from the isostructural substrate, with the ideal checkerboard-like cation ordering of a pyrochlore viewed along $\langle1\overline{1}0\rangle$ (Figure \ref{StructandChar}c and Figure \ref{suppCheck}). The film exhibits minimal cation disorder and no extended defects, such as antiphase boundaries that are common to pyrochlore thin films on YSZ substrates \cite{anderson2024defect}. Atomic resolution electron energy loss spectroscopy (EELS) mapping of elemental concentration further presents a well-ordered pyrochlore film atop a sharp substrate-film interface with at most one monolayer of interdiffusion (Figure \ref{StructandChar}d and Figure \ref{suppSTEM-EELS}). Both EELS and x-ray absorption spectroscopy (XAS) measurements of the V-$L_{2,3}$ edge show the expected peak shape for V$^{4+}$ in a nearly octahedral coordination environment.  


Based on the lattice mismatch between Y$_2$V$_2$O$_7$ and Y$_2$Ti$_2$O$_7$, fully strained films experience approximately 0.9\% tensile strain. Reciprocal space mapping shows that our Y$_2$V$_2$O$_7$ films are fully strained up to at least 45 atomic layers (12.8 nm). Beyond this critical thickness, films show partial relaxation indicated by a diffuse background spread towards the bulk Y$_2$V$_2$O$_7$ peak position (Figure \ref{suppRSMthick}). 



To probe the magnetic properties at the ultrathin limit of Y$_2$V$_2$O$_7$, we leverage the atomic layer precision of MBE to synthesize a series of (Y$_2$Ti$_2$O$_7$)$_m$/(Y$_2$V$_2$O$_7$)$_n$ superlattices where $m,n \in \mathbb{N}$ are the number of atomic planes in each repeat. Superlattices enable the characterization of the magnetic properties of the film in the ultrathin limit by maintaining the low dimensionality while increasing the total magnetic signal produced by multiple non-interacting repeats. While these samples allow us to avoid sample volume limits for our magnetic measurements, some prior studies have suggested that interlayer coupling can have an effect on magnetic properties \cite{cui2020correlation,liu2012metal}. We thus label each data point in magnetic data that arises from superlattice samples. In each superlattice, we hold $m$ = 6 so that each nonmagnetic Y$_2$Ti$_2$O$_7$ is a full unit cell thick to ensure negligible interlayer coupling between Y$_2$V$_2$O$_7$ layers: the magnetic exchange interaction is known to be vanishingly small beyond the nearest neighbors \cite{mena2014spin}. We vary $n \in \{2,6,10\}$ to compare directly with the thickness series and smoothly probe unit-cell and single tetrahedra thicknesses (Figure \ref{suppSLschem}). We adjust the number of superlattice repeats to maintain 30 total atomic layers of Y$_2$V$_2$O$_7$ to ensure a measurable magnetic signal. To confirm our measurements are not dominated by the effects of interdiffusion, we further synthesized a set of fully disordered Y$_2$Ti$_x$V$_{2-x}$O$_7$ samples with the same Ti:V ratio and overall thickness.

In situ RHEED shows the superlattices are smooth and highly crystalline, similar to single phase Y$_2$V$_2$O$_7$ or Y$_2$Ti$_2$O$_7$ films. Due to the similar atomic weight of titanium and vanadium, the superlattice layering is challenging to characterize with diffraction techniques (Figure \ref{suppSLxrd}). Only the (Y$_2$Ti$_2$O$_7$)$_6$/(Y$_2$V$_2$O$_7$)$_6$ superlattice shows a single weak XRD satellite peak about the 222 film peak near the expected superlattice peak position. This peak is absent in the corresponding Y$_2$TiVO$_7$ film. The low noise floor of synchrotron XRD more clearly reveals this presumed superlattice peak and confirms the absence of defect phases (Figure \ref{suppCHESS}). STEM imaging shows the superlattice forms the pyrochlore structure with minimal $A$-$B$ site cation disorder (Figure \ref{suppSL-STEM}) while atomic resolution EELS maps of vanadium and titanium concentration show increasing interdiffusion deeper in the film (Figure \ref{suppSL-EELS}).

\subsection{Magnetic Properties}

\begin{figure*}
    \centering
    \includegraphics[width=0.8\textwidth]{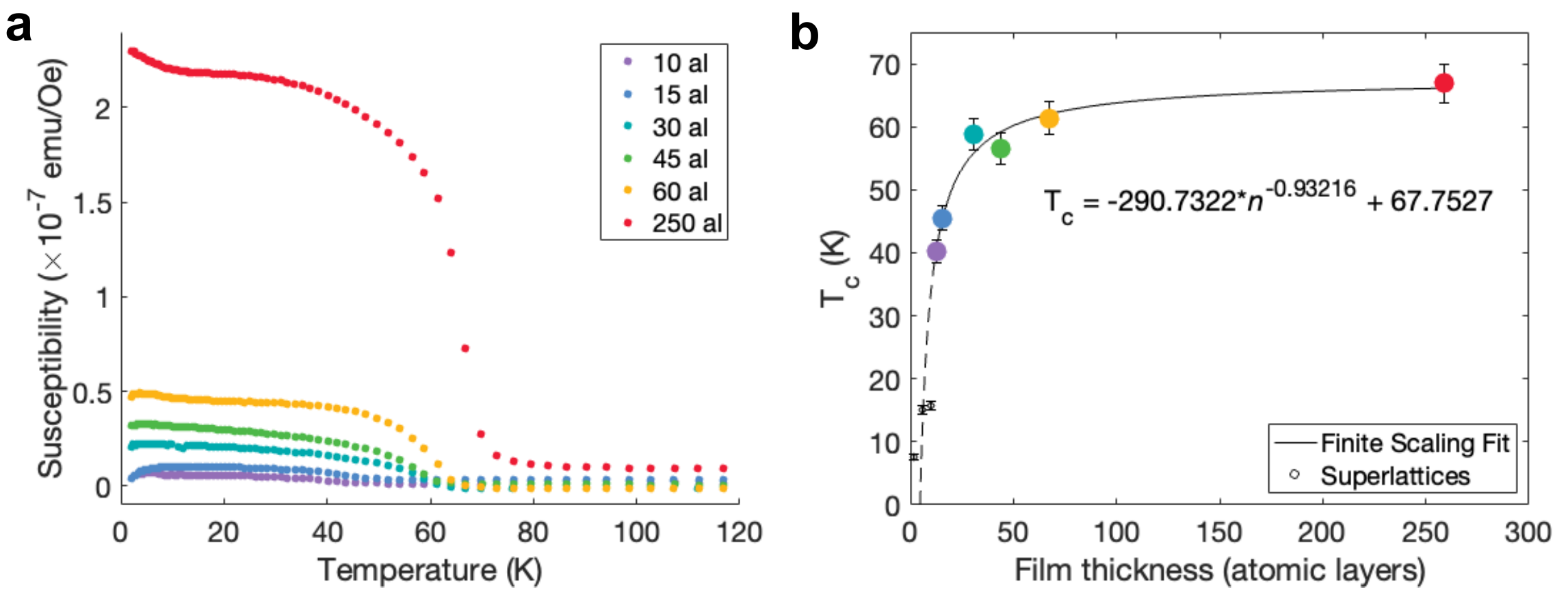}
    \caption{Magnetic susceptibility of Y$_2$V$_2$O$_7$ thickness series. \textbf{a} Field-cooled magnetic susceptibility measured in a field of 500 Oe perpendicular to $\langle111\rangle$ for each sample in the thickness series \textbf{b} Ferromagnetic transition temperature vs film thickness ($n$, atomic layers) with a finite-size scaling fit (solid line, extrapolated along dotted line) that shows excellent agreement with both thickness series (filled circles) and superlattice samples (open circles)}
    \label{SuscandTc}
\end{figure*}

All films are ferromagnetic insulators like bulk Y$_2$V$_2$O$_7$ and Lu$_2$V$_2$O$_7$. Y$_2$V$_2$O$_7$ films show a high surface resistance that increases in thinner films. Resistivity versus temperature shows fully insulating behavior (Figure \ref{suppRvsT}). Magnetic susceptibility measurements of the thickest Y$_2$V$_2$O$_7$ film in a 500 Oe in-plane applied field reproduces the bulk crystal susceptibility behavior with a sharp ferromagnetic transition at T$_\text{c}$ = 67 $\pm$ 3 K and a plateau at low temperatures (Figure \ref{SuscandTc}a, red). The field-cooled and zero-field curves diverge below T$_\text{c}$ (Figure \ref{suppSuscFC-ZFC}). Fitting the high temperature inverse susceptibility to the Curie-Weiss law yields $\theta_\text{CW} \approx$ 69 K and $\mu_\text{eff} \approx$ 1.83 $\mu_B$/V atom (Figure \ref{suppInvSusc}). For thinner films, the qualitative susceptibility behavior remains the same (Figure \ref{SuscandTc}a and Figure \ref{suppExtraSusc}) however the ferromagnetic T$_\text{c}$ decreases following the finite-size effect: 

\begin{equation}
\text{T}_\text{c}(n) = An^{-B} + C, 
\end{equation}

\noindent where $B = 1$ for a simple mean-field theory and $C$ is the bulk Curie temperature \cite{vaz2008magnetism}. The fit of this function to the measured T$_\text{c}$ of Y$_2$V$_2$O$_7$ films with thicknesses ranging from about 10-250 atomic layers (Figure \ref{SuscandTc}b) yields a scaling factor $B$ = 0.93, similar to that expected from mean-field theory, and $C$ = 67.7 K, consistent with the bulk Curie temperature T$_\text{c,bulk}$ = 68 K \cite{haghighirad2008powder,de2025observation}. The superlattice transition temperatures also align well with an extrapolation of the finite-size fit, suggesting that T$_\text{c}$ approaches 0 K in the monolayer limit, in contrast to theoretical predictions that the ferromagnetic transition temperature may be enhanced to room temperature or above in few monolayer films of the pyrochlore vanadates (Figure \ref{suppSLSusc}) \cite{pereiro2014topological}. 

X-ray magnetic circular dichroism measurements at the V-$L_{2,3}$ edge of a 60 atomic layer Y$_2$V$_2$O$_7$ film begins to show a signal near the ferromagnetic transition temperature (T$_\text{c} \sim$ 61 K) of susceptibility measurements (Figure \ref{suppXMCD}). This confirms that the ferromagnetism arises from the V$^{4+}$ ions, as expected.

\begin{figure*}
    \centering
    \includegraphics[width=0.8\textwidth]{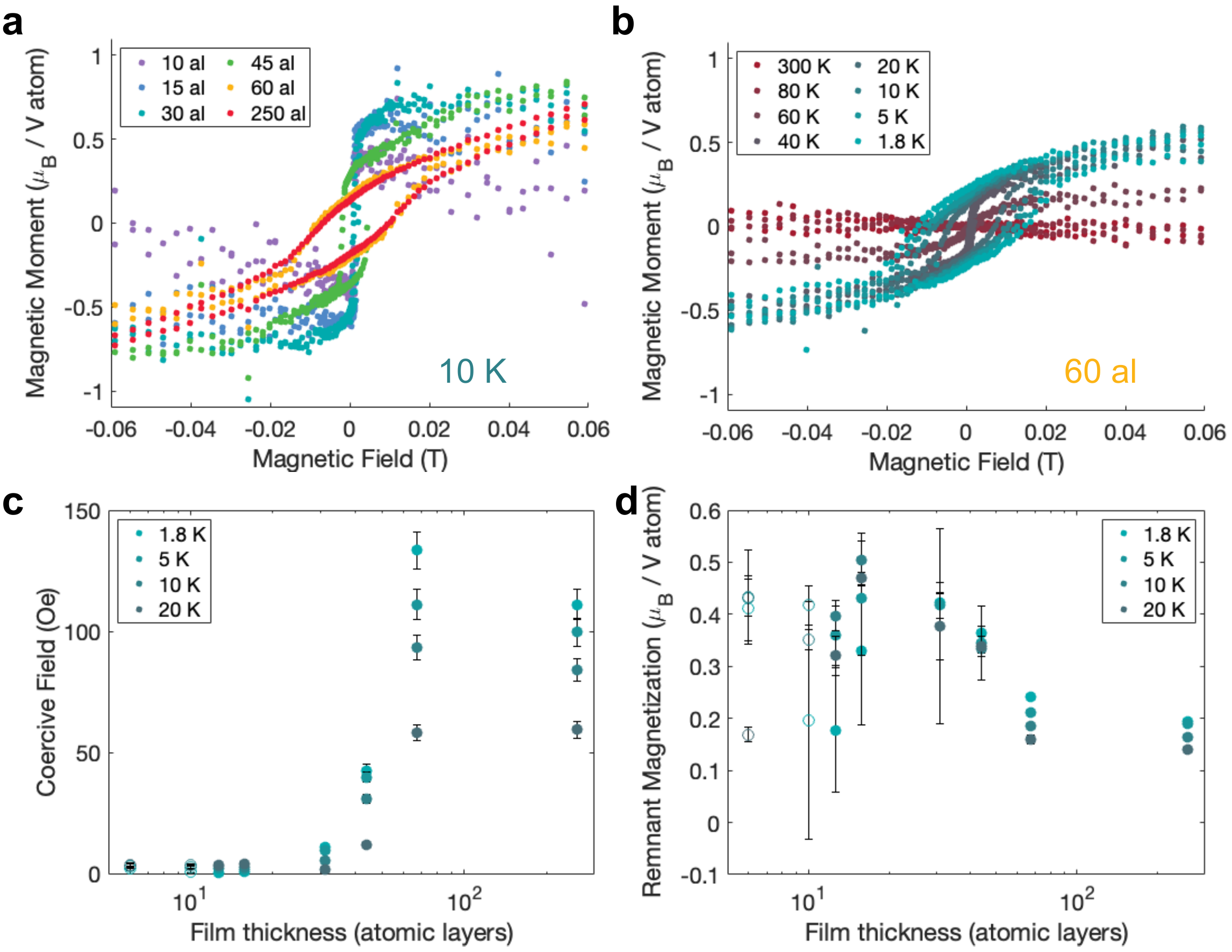}
    \caption{Magnetization vs. applied field loops. \textbf{a} Thickness dependence of magnetization vs. applied field loops measured at 10 K with field perpendicular to $\langle111\rangle$ \textbf{b} Temperature dependence of magnetization vs. applied field loops with field perpendicular to $\langle111\rangle$ for 60 atomic layers of Y$_2$V$_2$O$_7$, which has T$_\text{c}$ = 63 K \textbf{c} Coercive field (width) of magnetization loops vs film thickness and temperature with single-phase films (filled circles) and superlattices (open circles) \textbf{d} Remnant magnetization of Y$_2$V$_2$O$_7$ thin films vs. thickness and temperature}
    \label{MHandHc}
\end{figure*}

As hinted by the splitting of the field-cooled and zero-field-cooled susceptibility below the transition temperature, the magnetization of the thickest (250 atomic layer) film shows hysteresis under an applied field (Figure \ref{MHandHc}a, red). The narrow coercive field (at most 133 $\pm$ 7 Oe at 1.8 K) confirms that thin films of Y$_2$V$_2$O$_7$ behave as soft ferromagnets like fully oxidized bulk crystals of both Lu$_2$V$_2$O$_7$ and Y$_2$V$_2$O$_7$, which lack measurable hysteresis entirely \cite{knoke2007synthesis, de2025observation}. The hysteresis persists largely unchanged in films of approximately 60 atomic layers (Figure \ref{MHandHc}a,b). In thinner films, the coercivity shrinks approaching the ultrathin limit (Figure \ref{MHandHc}c and Figure \ref{suppSuppMH}). In magnetic thin films, coercivity tends to increase in the thinnest films as increasing strain induces defects, which act as pinning sites for magnetic domain walls \cite{ristau1999relationship}. Here, the growth in coercivity accompanies partial relaxation as film thickness increases past the critical value determined with RSM. This suggests that the 0.9\% tensile strain induced by the slightly larger Y$_2$Ti$_2$O$_7$ stabilizes the defect-free pyrochlore phase while relaxation leads to the formation of defects. A similar effect occurs in Ni/Cu thin films as the nickel thickness is increased past the critical value, leading to relaxation \cite{o1996anomalous}. In all films, the saturation magnetization is on the order of 1 $\mu_\text{B}$ per vanadium atom as expected for a collinear $d^1$ ferromagnet. The small volumes of the films lead to sizable uncertainty in the conversion to $\mu_\text{B}$/V, which complicates precise interpretation of the saturation value and direct comparison between films. However, the consistent magnitude throughout the thickness series suggests film quality does not degrade significantly down to at least 15 atomic layers. The $n = 10$ superlattice exhibits a larger saturated magnetization than the 10 atomic layer Y$_2$V$_2$O$_7$ film, indicating the high quality of the vanadate layers within the superlattice (Figure \ref{suppSLMH}). As $n$ decreases, the magnetic moment decreases until no signal is measured in the $n = 2$ film even at 1.8 K, as expected from the vanishing T$_\text{c}$.

\begin{figure*}
    \centering
    \includegraphics[width=\textwidth]{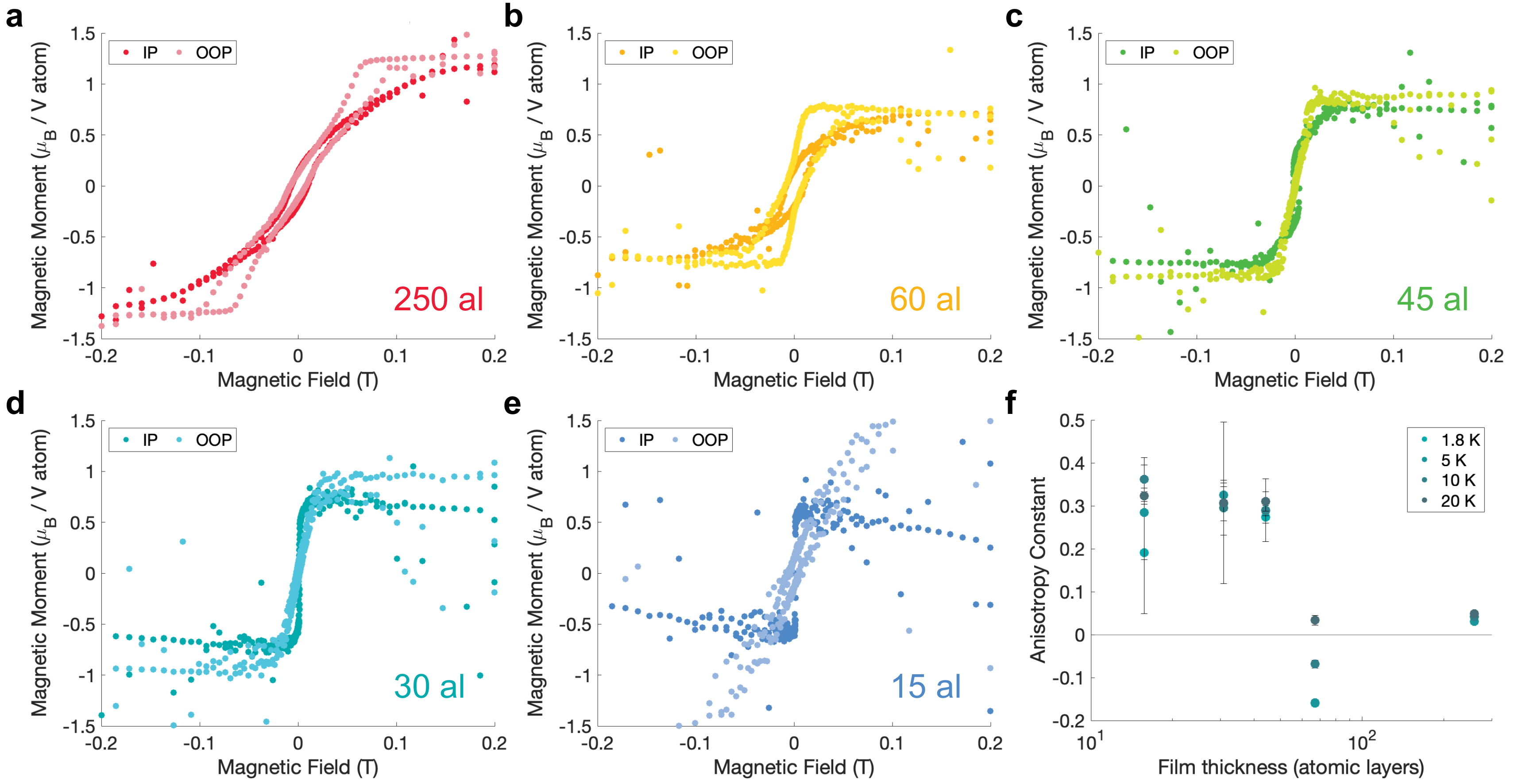}
    \caption{Magnetic anisotropy of Y$_2$V$_2$O$_7$ thin films. \textbf{a-e} Magnetization vs. applied field loops at 10 K for a field applied within the plane of the film (IP, perpendicular to $\langle111\rangle$) and out-of-plane (OOP, field along $\langle111\rangle$) for films with thickness (a) 250, (b) 60, (c) 45, (d) 30, and (e) 15 atomic layers \textbf{f} The anisotropy constant $\Delta$M$_\text{r}$ = M$_{\text{r,IP}}$ - M$_{\text{r,OOP}}$, assuming constant 1 $\mu_\text{B}$ saturation, showing a change in anisotropy between the 45 and 60 atomic layer films}
    \label{aniso}
\end{figure*}

Comparing the magnetization in a field applied within the plane of the film (perpendicular to $\langle111\rangle$) and out-of-plane (along $\langle111\rangle$), the magnetically easier axis transitions from out-of-plane in the thickest films to in-plane for films below 45 atomic layers thick (Figure \ref{aniso}a-e). This transition is coincident with both the onset of partial strain relaxation and the development of measurable ferromagnetic hysteresis. Accordingly, in films that exhibit this partial relaxation (60 and 250 atomic layers, Figure \ref{aniso}a,b), the out-of-plane magnetization conforms to the in-plane behavior at low fields. This indicates that the relaxed portion of the film may have distinct anisotropy. This change in anisotropy in partially relaxed films can also be seen in XMCD signal measured at normal and grazing incidences (Figures \ref{suppAngleXAS}-\ref{suppXMCDHystersis}). A change in easy axis represents a change of anisotropy, which can be quantified by an anisotropy constant defined as $\Delta \text{M}_r/\text{M}_s = (\text{M}_\text{r,IP}-\text{M}_\text{r,OOP})/\text{M}_s$. Here, due to the uncertainty in the exact value of the magnetic moment between films, we take M$_s$ to be the bulk value, 1 $\mu_B$/V$^{4+}$ ion. Quantifying the anisotropy in our films, we see a transition from positive (in-plane easy axis) towards negative (out-of-plane) anisotropy between the 45 and 60 atomic layer films (Figure \ref{aniso}f), as expected. Full characterization of magnetic behavior of the Y$_2$V$_2$O$_7$ thickness series in an out-of-plane field is presented in supplementary figure \ref{suppSuppMH-OOP}-\ref{suppTcThickOOP}. The $n = 10$ superlattice shows a strong in-plane easy axis (Figure \ref{suppSLOOP}). The anisotropy is less dramatic in the $n = 6$ superlattice, possibly due to the effects of interdiffusion. 

\section{Discussion and Conclusion}
We explore magnetism in ultrathin films of the pyrochlore ferromagnetic insulator Y$_2$V$_2$O$_7$. An important step towards the development of low-power magnonic devices, we synthesize the first reported vanadate pyrochlore thin films. While films on isostructural Y$_2$Ti$_2$O$_7$ substrates readily form the pyrochlore phase, YVO$_x$ deposited on commercial YSZ(111), a popular substrate for thin film pyrochlore synthesis, forms a distinct defect phase, which lacks the kagome vanadium planes that lead to nontrivial magnon topology. Y$_2$Ti$_2$O$_7$, in contrast, forms the pyrochlore phase on YSZ(111) and may be used as a buffer layer to seed pyrochlore Y$_2$V$_2$O$_7$ films. Based on structural characterization with in situ RHEED, XRD, and HAADF-STEM imaging, our films on Y$_2$Ti$_2$O$_7$ substrates are higher quality than most bulk crystal Y$_2$V$_2$O$_7$ \cite{haghighirad2008powder,haghighirad2008crystal}. The films, especially those that are fully strained, are smooth, highly crystalline, and single phase with very little cation disorder. (Y$_2$Ti$_2$O$_7$)$_6$/(Y$_2$V$_2$O$_7$)$_n$ superlattices are similarly high quality with respect to to strain and pyrochlore structure. However, marked interlayer cation diffusion within these samples may affect their magnetism, and although they display higher ferromagnetic ordering temperatures than their fully disordered counterparts (Figure \ref{suppSLSusc}), further superlattice studies are needed to determine whether these samples can fully allow us to probe magnetism down to a two atomic layer ($n$ = 2) limit.  

The magnetic properties of the thickest film (88 nm, approximately 250 atomic layers) largely reproduce the behavior of bulk crystals. A Curie-Weiss fit of the inverse susceptibility yields a $\mu_\text{eff}$ of 1.83 $\mu_B$ per vanadium ion, which is fairly consistent with the theoretical free ion value, 1.73 $\mu_\text{eff}$, and that found from fitting the inverse susceptibility of Lu$_2$V$_2$O$_7$ bulk crystals, 1.75 to 1.89 $\mu_\text{eff}$ per vanadium \cite{zhou2008magnetic,shamoto2001light,su2019asymmetric,haghighirad2008powder}. The overestimation of this value derives from spin orbit coupling \cite{su2019asymmetric}. Our fit also gives a $\theta_\text{CW}$ of 69 K, consistent with the ferromagnetic transition temperature determined by the maximum rate of change of the susceptibility. Intriguingly, all Curie-Weiss fits to bulk crystal Lu$_2$V$_2$O$_7$ data yield a $\theta_\text{CW}$ about 30 K higher than the true Curie temperature and show a deviation from linear Curie-Weiss behavior at 130 -- 170 K, which is attributed to the formation of ferromagnetic clusters or magnetic polarons above the true ferromagnetic transition \cite{zhou2008magnetic,shamoto2001light,su2019asymmetric,knoke2007synthesis,biswas2013crystal}. The Y$_2$V$_2$O$_7$ thin film, in contrast, conforms well to linear Curie-Weiss behavior down to the ferromagnetic transition temperature, which suggests the formation of ferromagnetic clusters is absent or suppressed. 

As film thickness decreases, the measured Curie temperature falls according to finite-size effects with a scaling factor similar to that expected from mean-field theory. Unfortunately, our results suggest that the T$_\text{c}$ of ultrathin pyrochlore vanadates will not approach room temperature as hoped \cite{pereiro2014topological}. However, further exploration of the pyrochlore-like defect phase that forms on YSZ (with T$_\text{c} \approx$ 140 K) may suggest pathways to enhance the ferromagnetic transition temperature in Y$_2$V$_2$O$_7$ films. Apart from the decay of T$_\text{c}$ with decreasing thickness, our films also exhibit magnetic hysteresis that manifests as the films begin to relax. The partial relaxation of films beyond the critical thickness may be accompanied by the formation of defects, which act as domain wall pinning sites. This behavior, also noted in Ni/Cu thin films \cite{o1996anomalous}, is contrary to the typical behavior of thin films that form defects to relieve increasing strain. 
We also find a consistent saturation magnetization of about 0.75 $\mu_B$ per vanadium atom for all films, which suggests that the crystal quality of the films is maintained throughout the thickness series down at least 15 atomic layers. 

Questions about magnetic anisotropy in the pyrochlore vanadates have attracted significant interest because of their potential impact on magnon transport. Disagreement about the strength of Dzyaloshinskii-Moriya (DM) interactions led to the suggestion that V$^{4+}$ in the pyrochlore structure possess significant single-ion anisotropy \cite{xiang2011single, biswas2013crystal}, contrary to the fact that such a term should be trivially proportional to identity in a quantum spin-1/2 system \cite{riedl2016ab}. Neutron inelastic scattering suggests that magnetic exchange is largely isotropic in Lu$_2$V$_2$O$_7$ \cite{mena2014spin}, though fits to magnetic susceptibility may indicate a change in anisotropy below 170 K \cite{biswas2013crystal} or the ferromagnetic Curie temperature \cite{su2019asymmetric}. Higher order anisotropic exchange is alternately determined to be non-negligible \cite{riedl2016ab} or dismissed as perturbatively small \cite{hickey2025order}. Regardless, anisotropy is critically important to the topological behavior of magnons in the pyrochlore vanadates \cite{miyahara2007orbital,su2017magnonic,mook2014magnon} and may be modified in thin films via strain, doping, or dimensionality \cite{su2017magnonic}.

In thin films, we see a marked transition of anisotropy that accompanies the onset of partial relaxation between films of about 45 and 60 atomic layers. In thicker films, the magnetically easier axis lies in the out-of-plane 111 direction. At about 45 atomic layers thick, the easy axis transitions to within the plane of the film, partial relaxation occurs, and hysteresis opens. This change in anisotropy is opposite of that expected based on shape anisotropy in the thin film geometry where the thinnest films should exhibit an out-of-plane easy axis to reduce demagnetizing effects \cite{spaldin2010magnetic}. Bulk Lu$_2$V$_2$O$_7$ hosts a 100 magnetic easy axis \cite{onose2010observation}, while theoretical studies suggest that individual V$^{4+}$ ions have an easy axis along 111 \cite{hickey2025order,xiang2011single,biswas2013crystal}. We see unique behavior that suggests strain may indeed modify magnetic anisotropy and therefore impact magnon topology in thin films of Y$_2$V$_2$O$_7$. The tensile strain may distort the nearly octahedral coordination environment about vanadium atoms, leading to a change in inter- and intralayer exchange interactions and thus the overall magnetic anisotropy \cite{su2017magnonic}. Future explorations of the magnetism of vanadate pyrochlore thin films under strain may open the door to tunable anisotropy and magnon topology and selection of ideal properties for fully dissipationless magnonic devices.

\section{Methods}\label{sec:methods}

The Y$_2$Ti$_2$O$_7$ substrate crystal growth was performed using a conventional RF-heated Czochralski setup equipped with a crucible balance. The crystal was grown by the top-seeded solution growth technique (TSSG) using a flux based on TiO$_2$ and BaB$_2$O$_4$ at an initial melt temperature of about 1600$^\circ$C. An iridium crucible (about 58 mm inner diameter) embedded in ZrO$_2$ and Al$_2$O$_3$ insulation was used. An actively heated iridium afterheater and a lid with an opening for the seed holder were placed on top of the crucible. The growth of the crystal occurred at a rate of 0.35 mm$\cdot$h$^{-1}$ using automatic diameter control and a rotation rate of 10 rpm. An Ar/O$_2$ gas mixture (0.93\% O$_2$) at atmospheric pressure was used as growth atmosphere. Substrates of various sizes were prepared by CrysTec GmbH (Berlin, Germany) from the grown crystals.




We synthesize pyrochlore Y$_2$V$_2$O$_7$ thin films and (Y$_2$Ti$_2$O$_7$)$_m$/(Y$_2$V$_2$O$_7$)$_n$ superlattices with reactive-oxide molecular beam epitaxy in a Riber C21 dual chamber system. Yttrium, titanium, and vanadium are evaporated from elemental sources with fluxes of approximately 10$^{12}$ $\frac{\text{atoms}}{\text{cm}^2\cdot \text{s}}$ roughly calibrated in situ with a quartz crystal microbalance (Figure \ref{suppAFMstoich} and \ref{suppXRDstoich}). YVO$_x$ deposited on (111)(ZrO$_2$)$_{0.905}$(Y$_2$O$_3$)$_{0.095}$ (yttria-stabilized zirconia or YSZ) forms a pyrochlore-like defect phase (Figure \ref{suppCharYVO_YSZ}-\ref{suppSTEMYVO_YSZ}). The true pyrochlore phase forms on noncommercial, isostructural Y$_2$Ti$_2$O$_7$(111) substrates at 800$^\circ$C in 2 $\times$ 10$^{6}$ torr O$_2$. Film crystallinity and surface morphology are monitored and optimized with in situ reflection high-energy electron diffraction (RHEED). 

Based on a monolayer deposition time inferred from RHEED intensity oscillations and ex-situ x-ray reflectivity measurements of calibration samples, we synthesize a series of pyrochlore Y$_2$V$_2$O$_7$ films simultaneously on YSZ(111) and Y$_2$Ti$_2$O$_7$(111) with thicknesses ranging from 10-260 atomic layers (1.67-43.33 unit cells). We first explore proof-of-concept pyrochlore superlattice growth of high atomic contrast (Y$_2$Ti$_2$O$_7$)$_m$/(Tb$_2$Ti$_2$O$_7$)$_n$ films on YSZ(111) (Figure \ref{suppTTOsl}). We then synthesize a set of superlattices with 6 atomic layer (1 unit cell) non-magnetic Y$_2$Ti$_2$O$_7$ spacer layers separating 2, 6, or 10 atomic layer Y$_2$V$_2$O$_7$ with a total of 30 atomic layers of Y$_2$V$_2$O$_7$ in each film. As a control, we produce submonolayer-shuttered disordered Y$_2$Ti$_x$V$_{2-x}$O$_7$ films with the same Ti:V ratio and overall thickness as each superlattice.

Using a Malvern Panalytical Empyrean four circle diffractometer (Cu K$\alpha_1$ radiation, $\lambda$ = 1.5406 \AA) equipped with a Ge(220)x2 monochrometer and Pixcel3D detector, we characterize the structural quality and strain state of our films with x-ray reflectivity (XRR), x-ray diffraction (XRD), and reciprocal space mapping (RSM). 

With an Asylum MFP-3d Origin+ atomic force microscope (Oxford Instruments) in tapping mode, we measure sample surface roughness and visualize its topography.

We prepare lamellae for scanning transmission electron microscopy with a Helios 660 focused ion beam system. To verify film structure and superlattice layering, we conduct high-angle annular dark field (HAADF) STEM imaging and electron energy loss spectroscopy (EELS) on a ThermoFisher Scientific Themis Z STEM operating at 200 kV with a convergence angle of 18.9 mrad and collection angle range of 64-200 mrad. Low-angle annular dark field imaging (LAADF) with collection angles from 23-128 mrad yield higher substrate-film contrast. For each micrograph, a stack of short exposure scans are cross-correlated and summed to correct for beam instabilities and drift. 

We measure the magnetic properties of the films on a Quantum Design Magnetic Property Measurement System 3 at Harvard's Laukien-Purcell Instrumentation Center. For measurements with the magnetic field applied within the plane of the film, samples are mounted on a quartz rod using GE Varnish (Oxford Instruments) within a clear plastic straw. For out of plane fields, the samples are wedged in place in a straw and secured with clear gel capsules. We measure the magnetic background of a Y$_2$Ti$_2$O$_7$ substrate annealed in the MBE at growth conditions. Samples are field-cooled in a 2 kOe applied field and susceptibility is measured in a 500 Oe field unless otherwise specified. Susceptibility measurements are corrected for the substrate background and the temperature independent diamagnetism, then normalized by volume to yield units of $\mu_\text{B}$/ V atom. Magnetization versus applied field curves are corrected with substrate subtraction and volume normalization. Ferromagnetic critical temperatures are defined as the temperature at the highest rate of change in susceptibility.

X-ray absorption spectroscopy (XAS) carried out at beamline 6.3.1 at the Advanced Light Source confirms the expected oxidation state and local symmetry environment of V$^{4+}$ in the Y$_2$V$_2$O$_7$ thin films. Further, we conduct x-ray magnetic circular dichroism measurements under $\pm$1.4 T magnetic field to verify the ferromagnetic behavior of the vanadium ions.

Further x-ray magnetic circular dichroism measurements were performed at the I10-1 beamline at the Diamond Light Source. Measurements were done on the V-$L_{2,3}$ edge at 20 K at grazing and normal incidences. For measurements at 1.5 T, we measured with both positive and negative circularly polarized light as well as at both positive and negative field. For measurements at 0.02 T, we measured with only positive field so as to probe the same magnetic state with each measurement. Field dependent data was taken at 514.08 eV with background taken at 510 eV. After background subtraction, field dependent data was normalized to the total of the background and on-peak data.

Synchrotron surface x-ray diffraction measurements are performed at the beamline sector 7-ID-C of the Advanced Photon Source, Argonne National Laboratory, using a six-circle Huber diffractometer in Psi-C geometry. The incident x-ray beam is monochromated to 17.5 keV ($\lambda = 0.70846$ \AA) using a Si(111) double-crystal monochromator ($\Delta$E/E $\approx$ 1 × 10$^{-4}$), and focused to a 30 $\mu$m (vertical) × 50 $\mu$m (horizontal) beam spot size using Kirkpatrick–Baez mirrors The total photon flux at the sample is approximately 3 × 10$^{12}$ photons/s. Specular (00$L$) Bragg rod measurements are collected with an Eiger2 X 500K area detector up to $L$ = 3.9 reciprocal lattice units (r.l.u.), with incident and exit angles matched. Raw 2D detector images are corrected for background scattering, geometric factors, and pixel response non-uniformity. Reciprocal lattice vector ($L$) is converted to total angular deflection (2$\theta$) using $L = (\frac{2c_0}{\lambda})\sin(\frac{2\theta}{2})$ where $c_0$ is the substrate lattice parameter (10.01 \AA \ for Y$_2$Ti$_2$O$_7$) and $\lambda$ is the x-ray wavelength (0.70846 \AA).

Electronic transport measurements were conducted in a Physical Property Measurement System (Quantum Design) with a Dynacool closed-loop helium compression system.

\section*{Acknowledgments}
The authors acknowledge useful conversations with Johanna Nordlander. This work was supported by the Air Force Research Laboratory, Project Grant FA95502110429.
J.A.M. acknowledges support from the Packard Foundation and the Gordon and Betty Moore Foundation’s EPiQS Initiative, grant GBMF6760. D.F.S. acknowledges support from the NSF Graduate Research Fellowship No. DE-SC0021925.
This research used resources of the Advanced Light Source, which is a DOE Office of Science User Facility under contract no. DE-AC02-05CH11231. Crystal growth work at Johns Hopkins University was supported by the Institute for Quantum Matter, an Energy Frontier Research Center funded by DOE, Office of Science, Basic Energy Sciences under Award No. DE-SC0019331. This research was also performed on APS beam time proposal No. 1009892 from the Advanced Photon Source, a U.S. Department of Energy (DOE) Office of Science user facility operated for the DOE Office of Science by Argonne National Laboratory under Contract No. DE-AC02-06CH11357. Sample preparation with focused ion beam was performed at Harvard University's Center for Nanoscale Systems, a member of the National Nanotechnology Coordinated Infrastructure Network, supported by the NSF under grant no. 2025158. Electron microscopy was performed at MIT.nano's Characterization.nano facility with support from Aubrey Penn. The authors thank D. Cui and A. Lowe at Harvard's Laukien-Purcell Instrumentation Center for their assistance with SQUID magnetometry systems.


\printendnotes


\newpage

\clearpage

\onecolumngrid


\setcounter{equation}{0}
\setcounter{figure}{0}
\setcounter{table}{0}
\setcounter{page}{1}
\setcounter{section}{0}
\makeatletter
\renewcommand{\theequation}{S\arabic{equation}}
\renewcommand{\thefigure}{S\arabic{figure}}
\renewcommand{\bibnumfmt}[1]{[S#1]}
\renewcommand{\thesection}{S\arabic{section}}
\makeatother



\huge
\section*{Supplemental Materials: The evolution of magnetism in a thin film pyrochlore ferromagnetic insulator}

\Large
Margaret A. Anderson$\dagger$, Megan E. Goh$\dagger$, Yang Zhang, Kyeong-Yoon Baek, Michael Schulze, Mario Brützam, Christoph Liebald, Chris Lygouras, Dan Ferenc Segedin, Aaron M. Day, Zubia Hasan, Donald A. Walko, Hua Zhou, Peter Bencok, Alpha T. N'Diaye, Charles M. Brooks, Ismail El Baggari, John T. Heron, S. M. Koopayeh, Daniel Rytz, Christo Guguschev, and Julia A. Mundy$^*$

\normalsize


\begin{center}
    \includegraphics[width=0.75\textwidth]{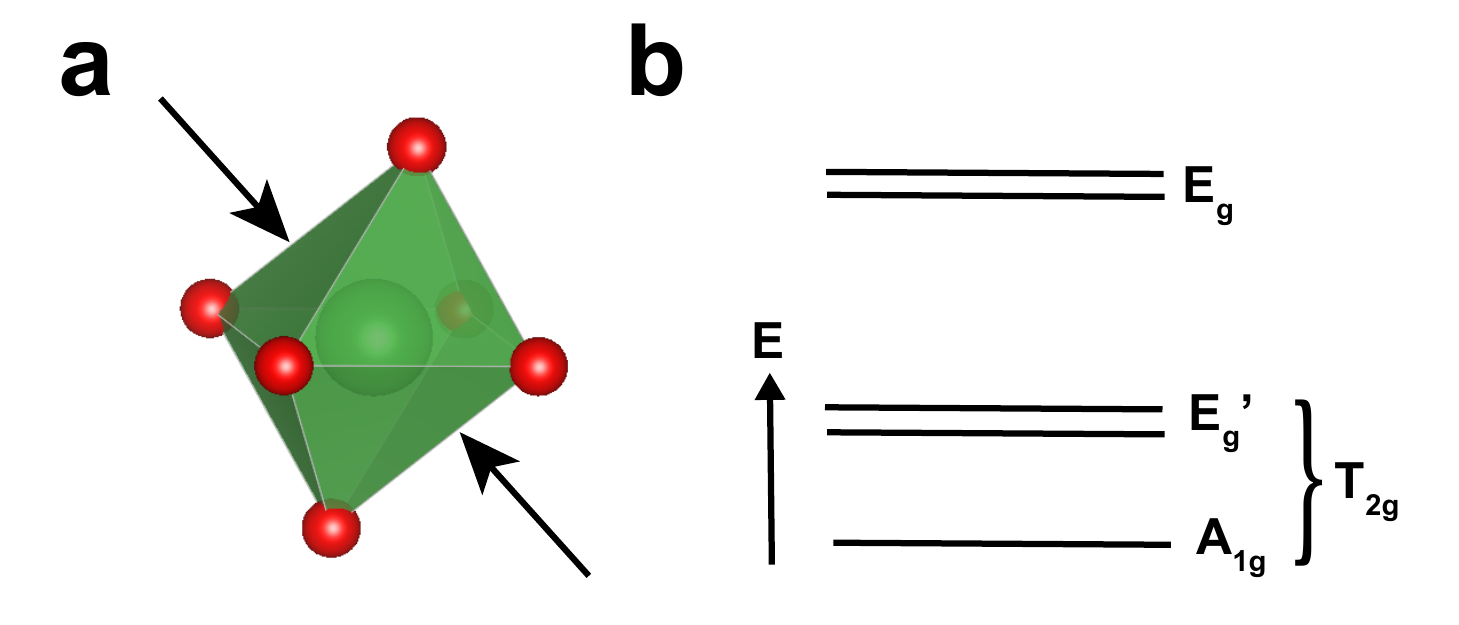}
\end{center}
    \captionof{figure}{Oxygen coordination and crystal field levels in Y$_2$V$_2$O$_7$. \textbf{a} The local oxygen coordination environment of V$^{4+}$ in Y$_2$V$_2$O$_7$, which is nearly octahedral with a slight trigonal distortion (emphasized with arrows) \textbf{b} The crystal electric field energy levels of V$^{4+}$ with configuration 3$d^1$ where the five 3$d$ orbitals are split into two degenerate upper E$_g$ levels and three lower energy T$_{2g}$ levels by the octahedral coordination; the three T$_{2g}$ states further split into one lowest lying A$_{1g}$ state and two higher energy E$_g$' levels due to the trigonal distortion}
    \label{suppTrigCF}

\begin{center}
    \includegraphics[width=\textwidth]{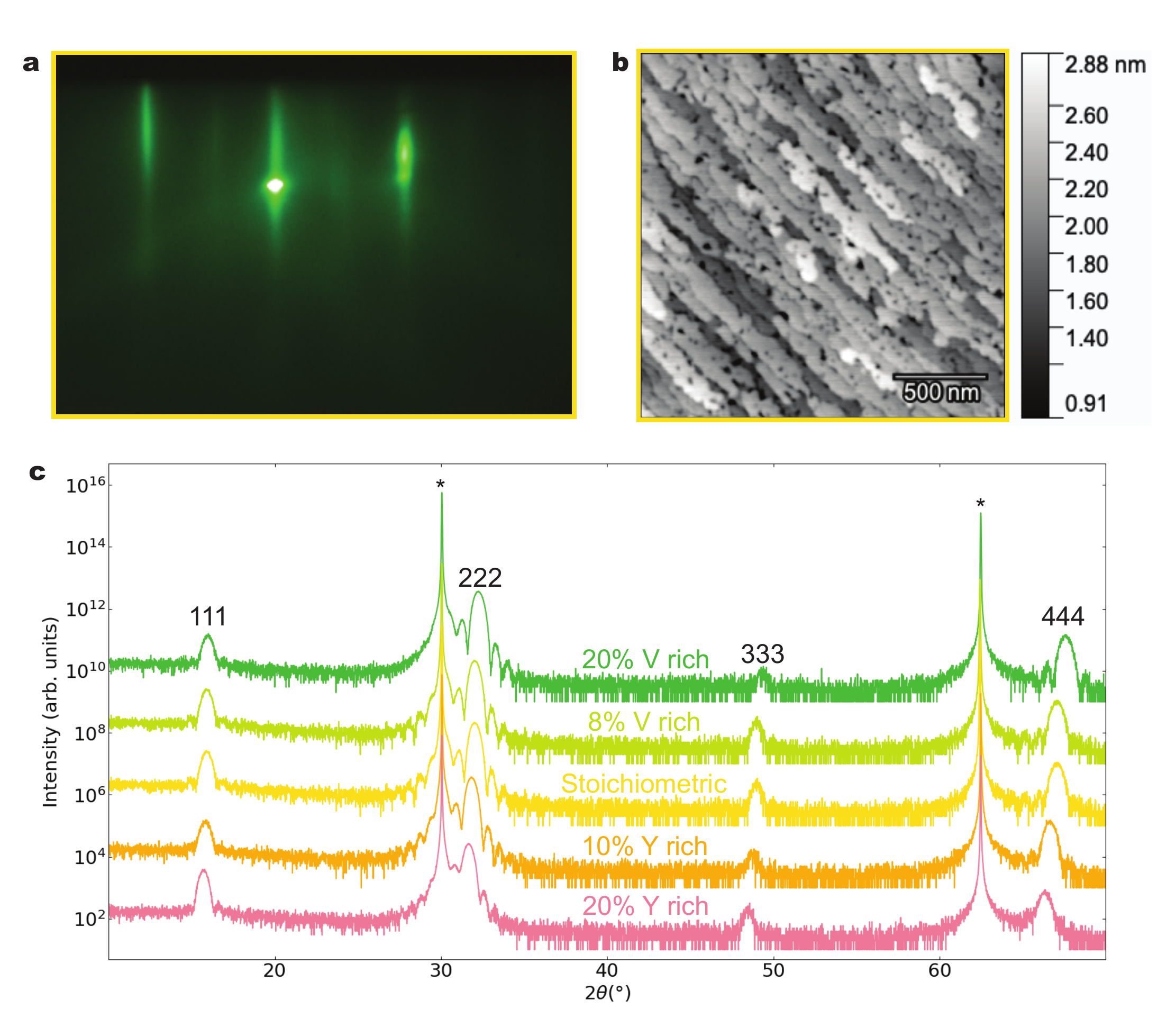}
\end{center}
    \captionof{figure}{Structural characterization of YVO$_x$ on YSZ(111). \textbf{a} Reflection high energy electron diffraction (RHEED) of YVO$_x$ on YSZ(111) \textbf{b} Atomic force microscopy of a film showing a smooth surface with an RMS roughness of 322 pm \textbf{c} X-ray diffraction of YVO$_x$ on YSZ(111) with varying stoichiometry. Asterisks indicate substrate diffraction peaks and lattice plane indices denote film peaks.}
    \label{suppCharYVO_YSZ}

\begin{center}
    \includegraphics[width=\textwidth]{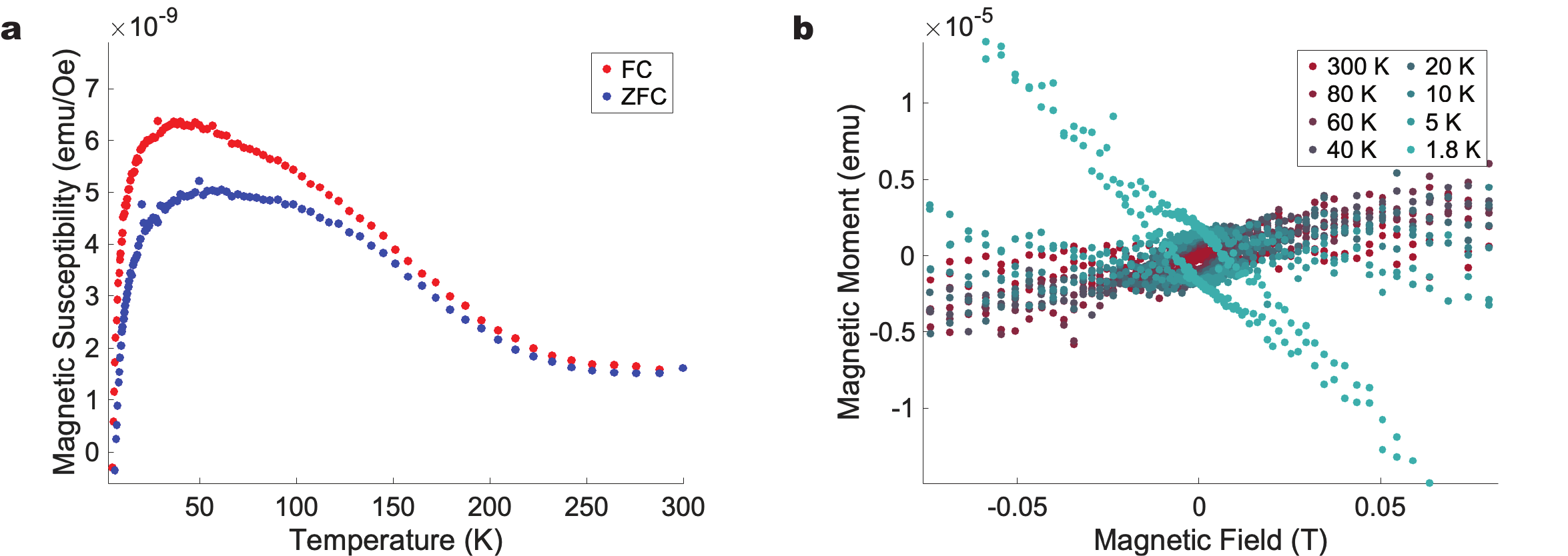}
\end{center}
    \captionof{figure}{Magnetic characterization of YVO$_x$ on YSZ(111). \textbf{a} Field-cooled vs. zero-field-cooled (ZFC) magnetic susceptibility of stoichiometric film \textbf{b} Magnetization vs. in-plane applied field loops for stoichiometric YVO$_x$ on YSZ(111)}
    \label{suppSquidYVO_YSZ}

\begin{center}
    \includegraphics[width=\textwidth]{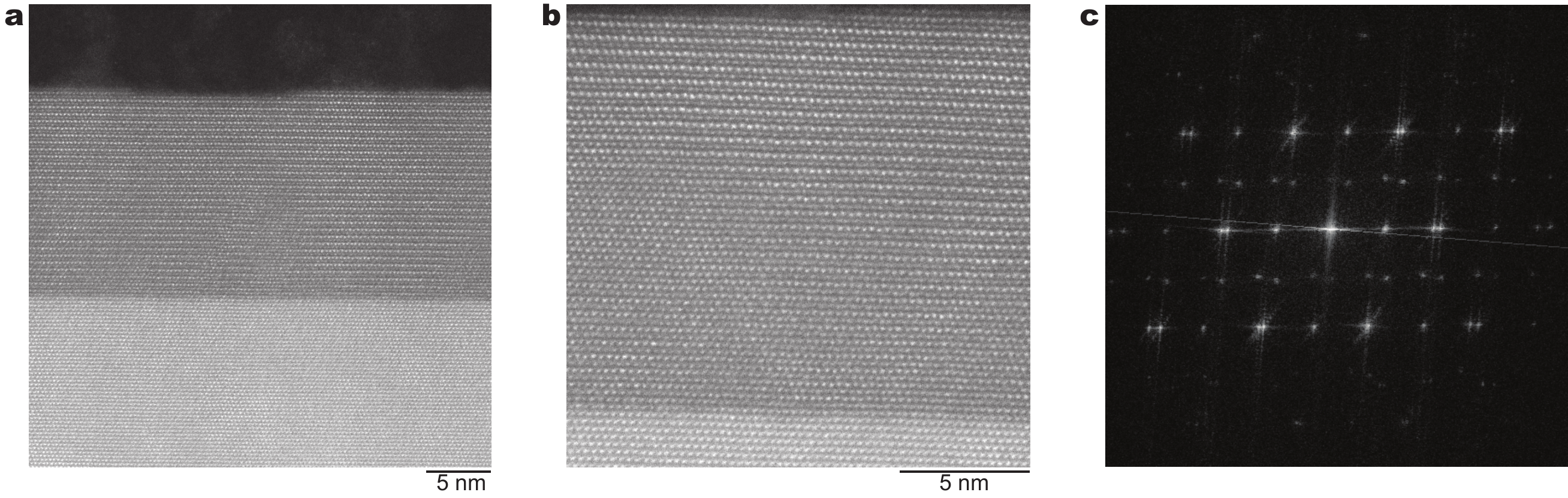}
\end{center}
    \captionof{figure}{STEM imaging of YVO$_x$ on YSZ(111). \textbf{a} a large field-of-view HAADF-STEM micrograph along YSZ $\langle1\overline{1}0\rangle$  \textbf{b} a higher magnification image of the film in (a) showing the lack of in-plane pyrochlore order in the film \textbf{c} a fast Fourier transform (FFT) of the YVO$_x$ film showing doubled intermediate peaks reminiscent of the pyrochlore-like defect fluorite structure, but with out-of-plane pyrochlore order}
    \label{suppSTEMYVO_YSZ}

\begin{center}
     \includegraphics[width=\textwidth]{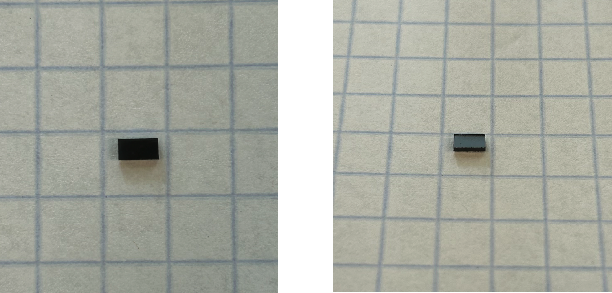}
\end{center}
    \captionof{figure}{Images of noncommercial Y$_2$Ti$_2$O$_7$ substrates. A top down view where each box is 5mm x 5mm (left); An angled view to show the smooth polished surface (right)}
    \label{suppYTOsub}

\begin{center}
    \includegraphics[width=\textwidth]{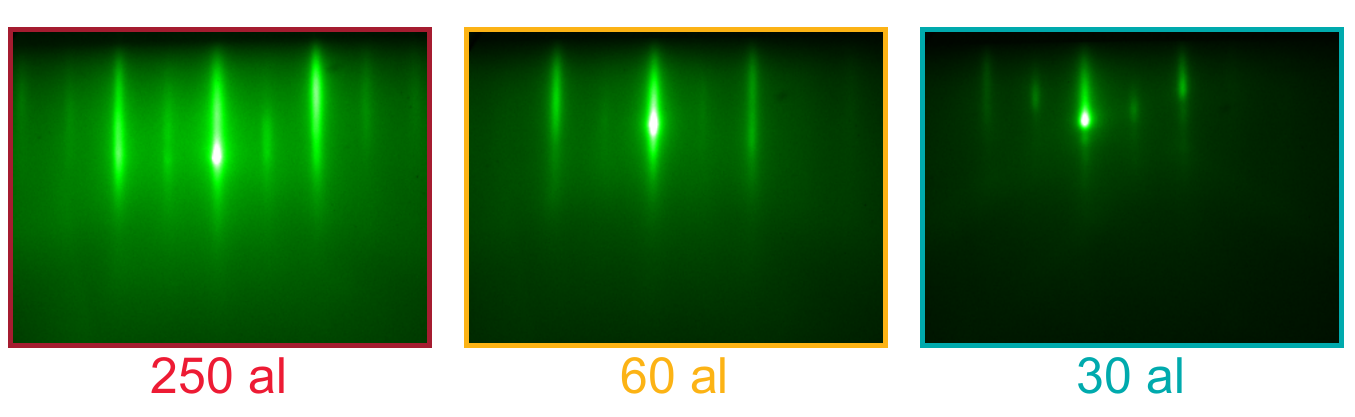}
\end{center}
    \captionof{figure}{Reflection high energy electron diffraction (RHEED) patterns after deposition for selected films in the Y$_2$V$_2$O$_7$ thickness series on Y$_2$Ti$_2$O$_7$.}
    \label{suppRHEED}


\newpage
\begin{center}
    \includegraphics[width=\textwidth]{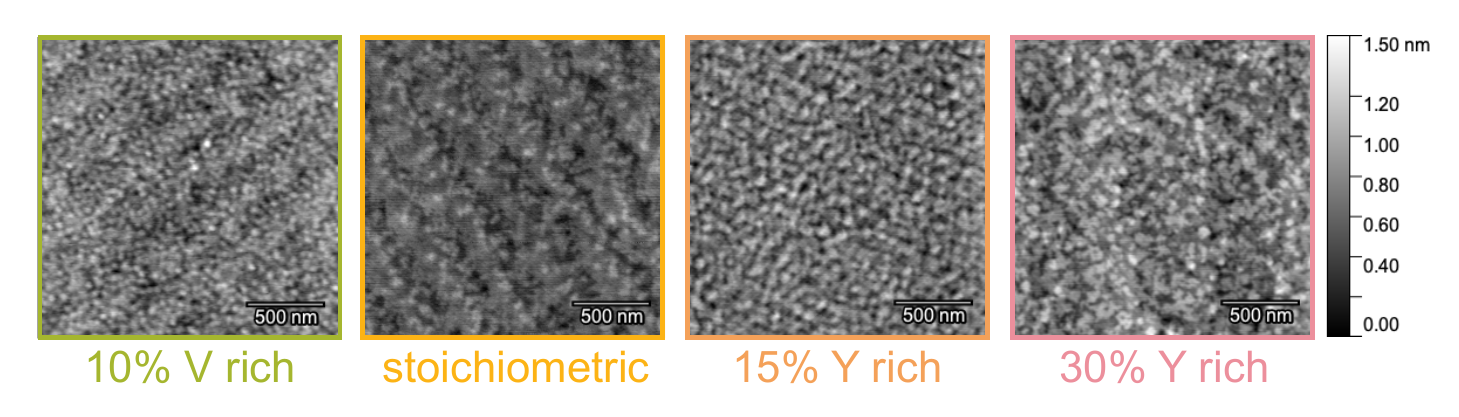}
\end{center}
    \captionof{figure}{Surface topography of Y$_2$V$_2$O$_7$ on Y$_2$Ti$_2$O$_7$ with varying stoichiometry. Atomic force microscopy shows smooth surfaces with a texture that depends on stoichiometry; RMS surface roughness values are 176.8 pm for the 10\% V rich film, 142.3 pm for the stoichiometric film, 190.8 pm for the 15\% Y rich film, and 208.8 pm for the 30\% Y rich film}
    \label{suppAFMstoich}

\begin{center}
    \includegraphics[width=\textwidth]{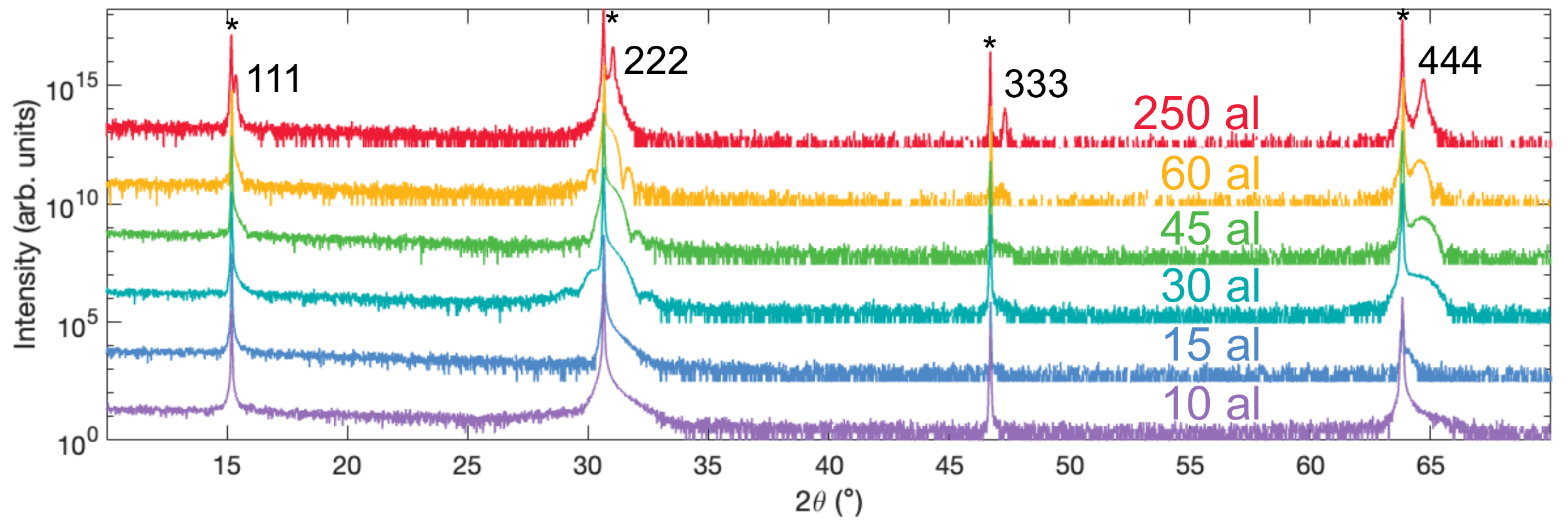}
\end{center}
    \captionof{figure}{Complete x-ray diffraction of the stoichiometric Y$_2$V$_2$O$_7$ thickness series on Y$_2$Ti$_2$O$_7$. Asterisks indicate substrate diffraction peaks and lattice plane indices denote film peaks.}
    \label{suppXRDfull}

\begin{center}
    \includegraphics[width=\textwidth]{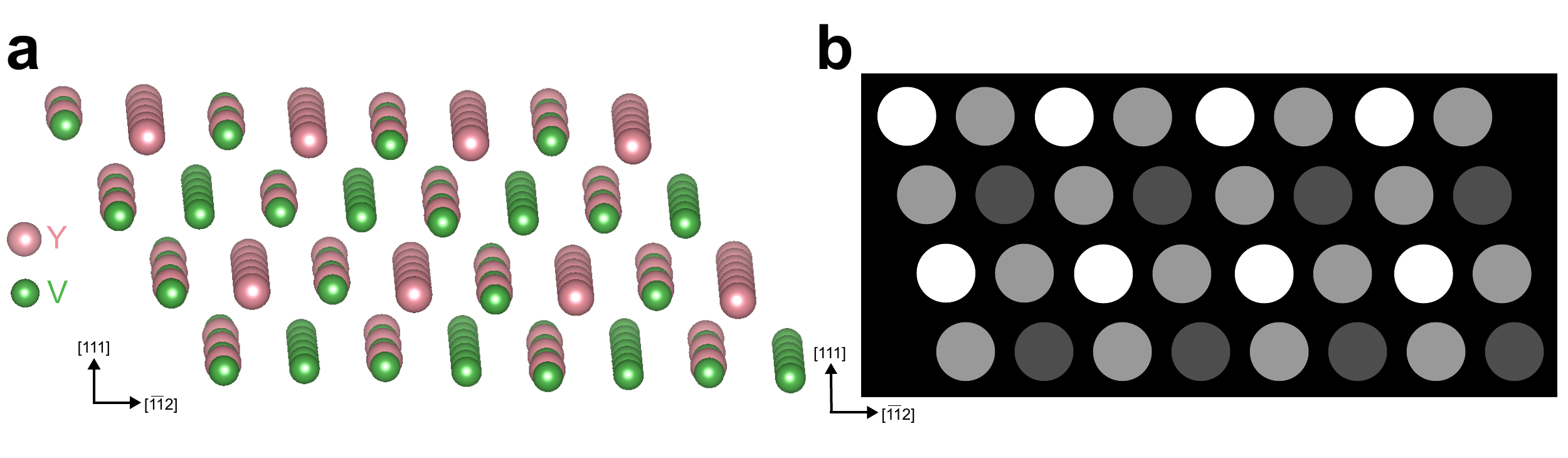}
\end{center}
    \captionof{figure}{An ideal pyrochlore in STEM imaging. \textbf{a} The cations in Y$_2$V$_2$O$_7$ viewed along $\langle 1\overline{1}0\rangle$ and slightly tipped off axis to show the composition of atomic columns; columns are either purely yttrium, purely vanadium, or alternating yttrium and vanadium out-of-plane \textbf{b} A schematic of the ideal pyrochlore structure along $\langle 1\overline{1}0\rangle$ in STEM where the apparent brightness of atomic columns scales with atomic number Z; alternating light and dark rows and slanted columns form a checkerboard pattern where the brightest quarter of spots are pure yttrium columns, the darkest quarter are pure vanadium columns, and half of the spots have an intermediate brightness corresponding to the mix of yttrium and vanadium}
    \label{suppCheck}

\begin{center}
    \includegraphics[width=\textwidth]{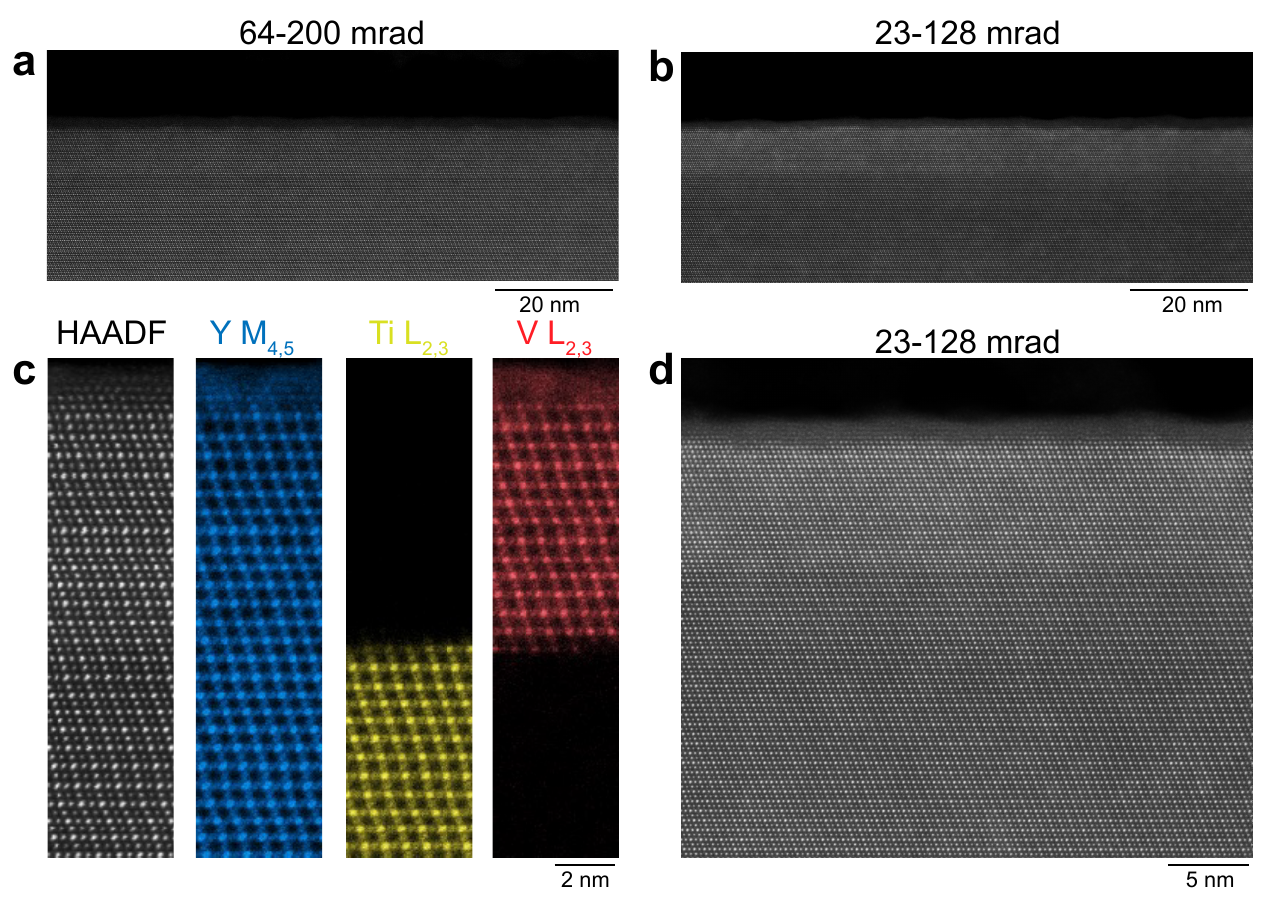}
\end{center}
    \captionof{figure}{Additional STEM-EELS imaging on Y$_2$V$_2$O$_7$ on Y$_2$Ti$_2$O$_7$. \textbf{a} A large field of view HAADF-STEM image with the same conditions as Figure \ref{StructandChar}c with collection angle range 64-200 mrad \textbf{b} A large field of view LAADF-STEM image with collection angle 23-128 mra, which emphasizes the substrate-film interface \textbf{c} From left to right: a HAADF-STEM micrograph of the region of interest, a concentration map of yttrium, a map of titanium concentration, and a vanadium concentration map of the same region as Figure \ref{StructandChar}d \textbf{d} A higher magnification view of (b)}
    \label{suppSTEM-EELS}

\newpage
\begin{center}
    \includegraphics[width=\textwidth]{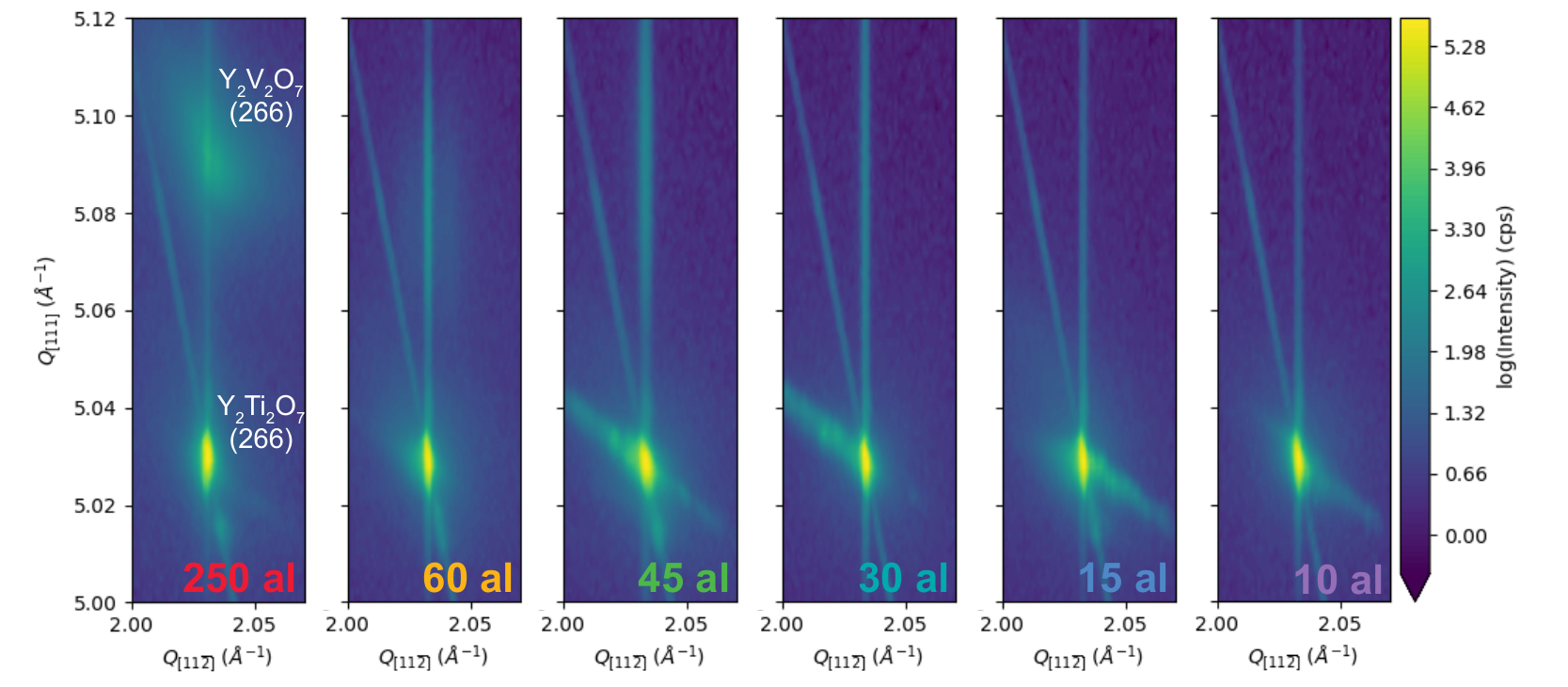}
\end{center}
    \captionof{figure}{Reciprocal space mapping of Y$_2$V$_2$O$_7$ thin films. The appearance of a diffuse background around the film peak for films with thickness above 45 atomic layers indicates partial strain relaxation.}
    \label{suppRSMthick}

\begin{center}
    \includegraphics[width=\textwidth]{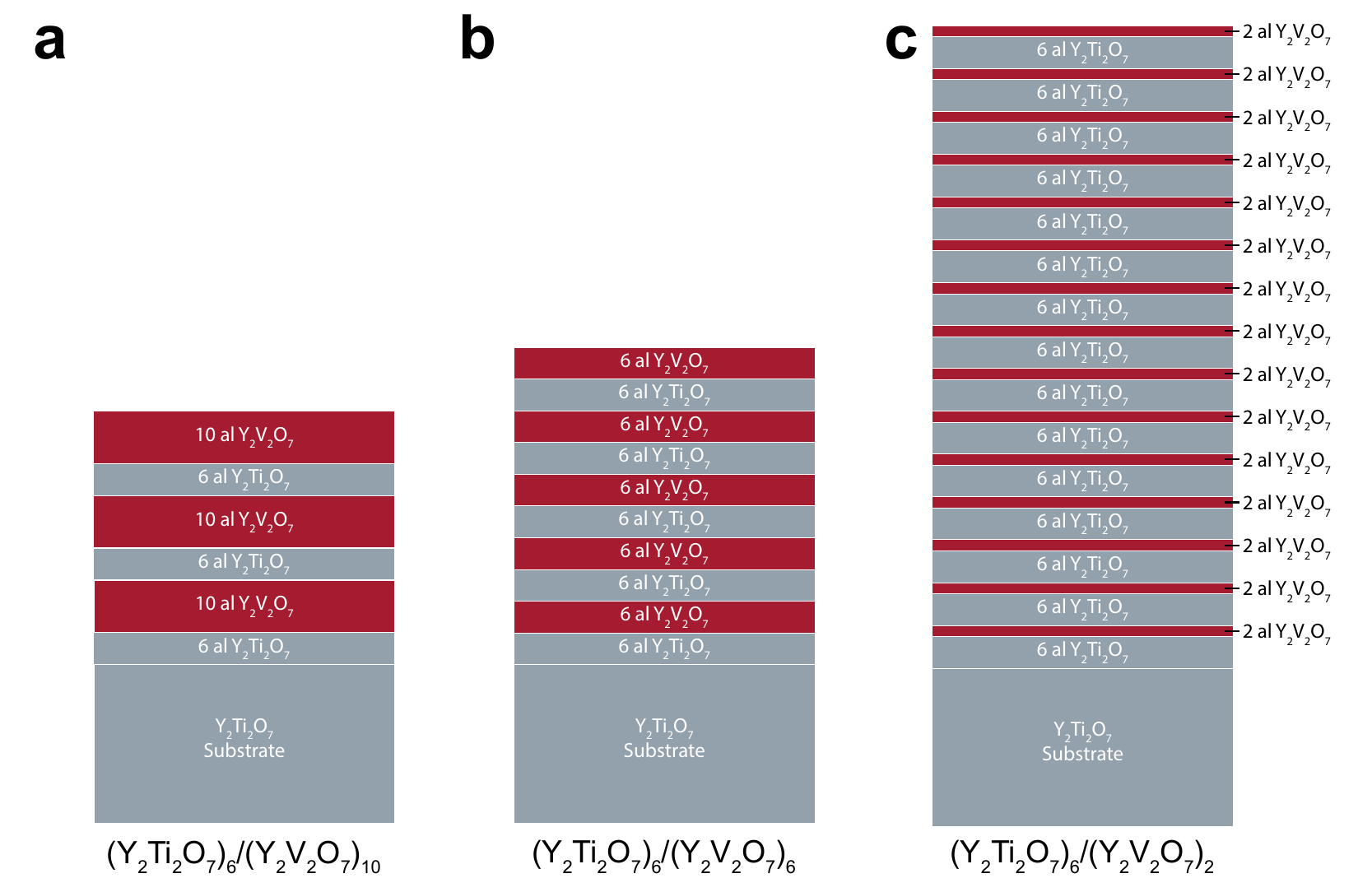}
\end{center}
    \captionof{figure}{Schematic of (Y$_2$Ti$_2$O$_7$)$_m$/(Y$_2$V$_2$O$_7$)$_n$ superlattices \textbf{a} The layering scheme for the (Y$_2$Ti$_2$O$_7$)$_6$/(Y$_2$V$_2$O$_7$)$_{10}$ superlattice with 3 repeats to yield a total of 30 atomic layers of Y$_2$V$_2$O$_7$ \textbf{b} The layering scheme for the (Y$_2$Ti$_2$O$_7$)$_6$/(Y$_2$V$_2$O$_7$)$_6$ superlattice with 5 repeats giving 30 total atomic layers of Y$_2$V$_2$O$_7$ \textbf{c} The layering scheme for the (Y$_2$Ti$_2$O$_7$)$_6$/(Y$_2$V$_2$O$_7$)$_2$ superlattice with 15 repeats to yield 30 total atomic layers of Y$_2$V$_2$O$_7$}
    \label{suppSLschem}

\begin{center}
    \includegraphics[width=\textwidth]{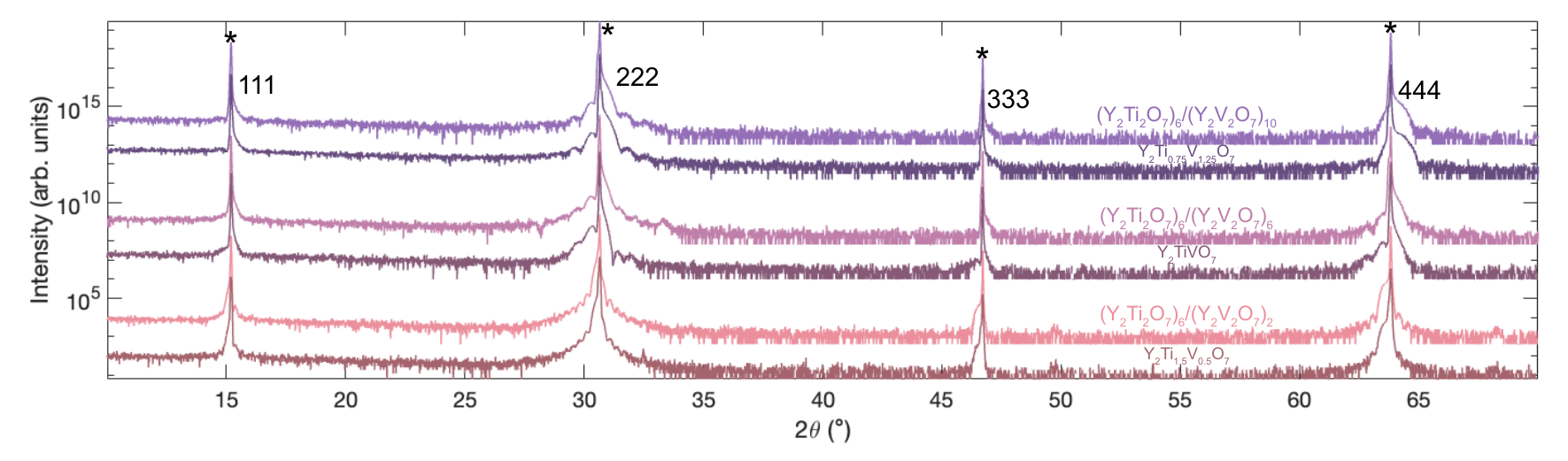}
\end{center}
    \captionof{figure}{X-ray diffraction of (Y$_2$Ti$_2$O$_7$)$_m$/(Y$_2$V$_2$O$_7$)$_n$ superlattices and their disordered counterparts. Y$_2$Ti$_2$O$_7$ substrate peaks indicated by asterisk; film peaks denoted by plane indices. Each pair of scans gives the ordered superlattice (top) corresponding to the schematics in Figure S12 and a corresponding solid solution (bottom) with the same composition and total thickness. A possible superlattice peak exists at $2\theta = 33^{\circ}$ within the (Y$_2$Ti$_2$O$_7$)$_6$/(Y$_2$V$_2$O$_7$)$_6$ superlattice, not seen within the corresponding disordered film Y$_2$TiVO$_7$.}
    \label{suppSLxrd}

\begin{center}
    \includegraphics[width=\textwidth]{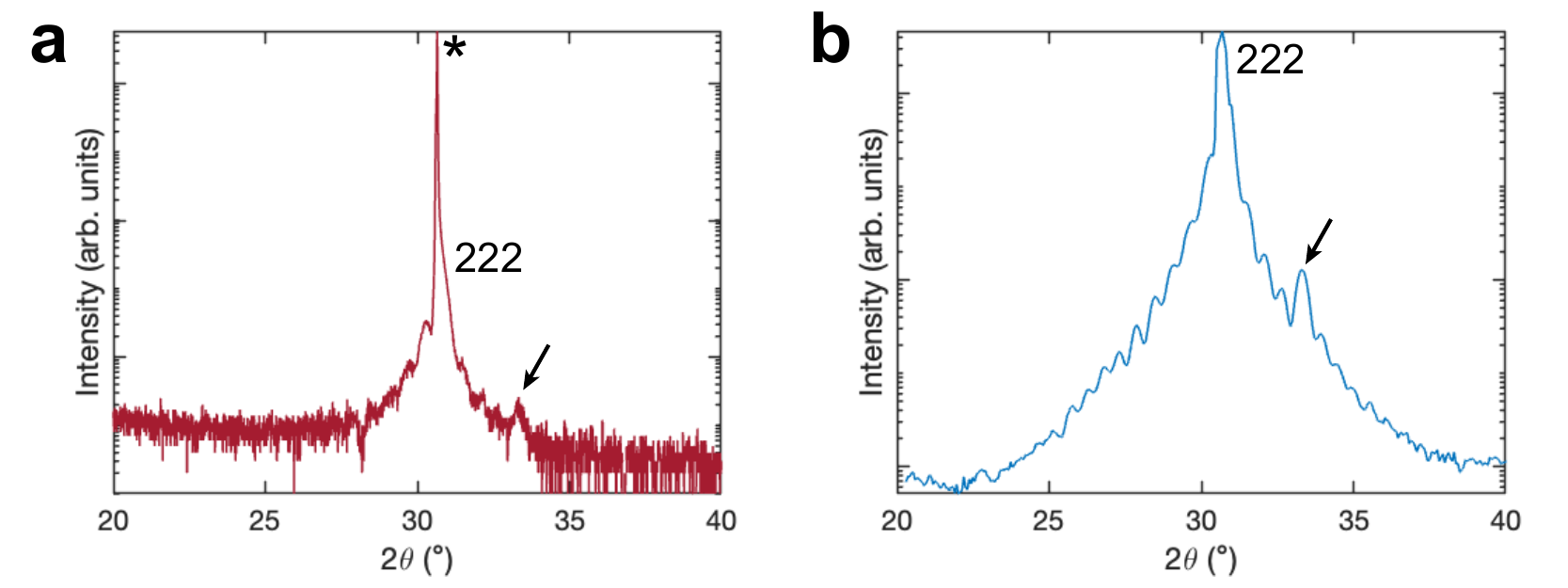}
\end{center}
    \captionof{figure}{X-ray diffraction of a (Y$_2$Ti$_2$O$_7$)$_6$/(Y$_2$V$_2$O$_7$)$_6$ superlattice. \textbf{a} X-ray diffraction of (Y$_2$Ti$_2$O$_7$)$_6$/(Y$_2$V$_2$O$_7$)$_6$ using a Malvern Panalytical Empyrean Cu K$\alpha$ diffractometer, which shows the 222 film peak adjacent to the 222 Y$_2$Ti$_2$O$_7$ substrate peak (indicated with an asterisk) with a possible weak superlattice peak (indicated by an arrow) \textbf{b} synchrotron x-ray diffraction data of the same sample as in (a) with the substrate subtracted showing the 222 film peak and a clear superlattice peak (indicated by an arrow)}
    \label{suppCHESS}

\begin{center}
    \includegraphics[width=\textwidth]{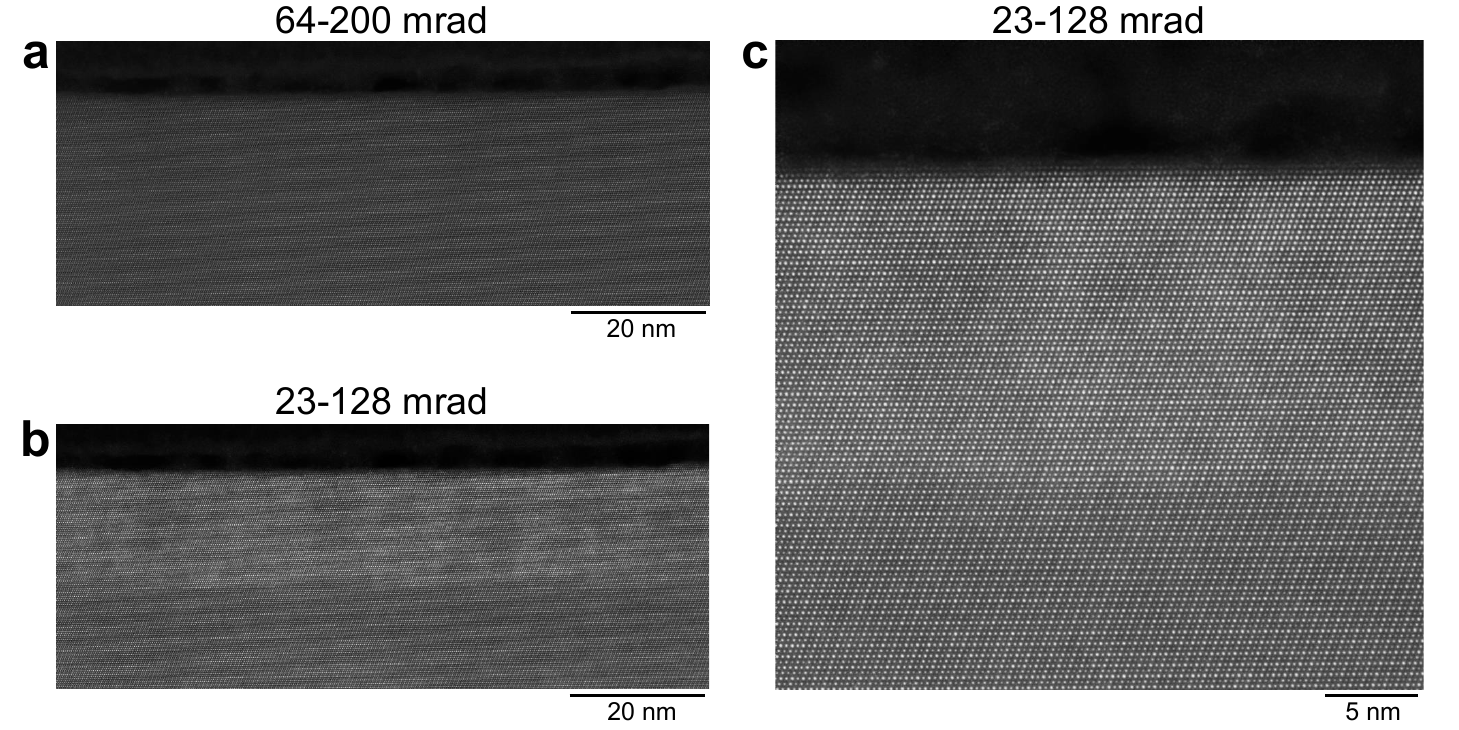}
\end{center}
    \captionof{figure}{STEM imaging of (Y$_2$Ti$_2$O$_7$)$_6$/(Y$_2$V$_2$O$_7$)$_6$ superlattice on Y$_2$Ti$_2$O$_7$. \textbf{a} A large field of view HAADF-STEM micrograph with collection angle range 64-200 mrad showing the high structural quality of the film over a large region \textbf{b} A large field of view LAADF-STEM image with collection angle 23-128 mrad, which highlights the substrate-film interface \textbf{c} A higher magnification view of (b)}
    \label{suppSL-STEM}

\begin{center}
    \includegraphics[width=\textwidth]{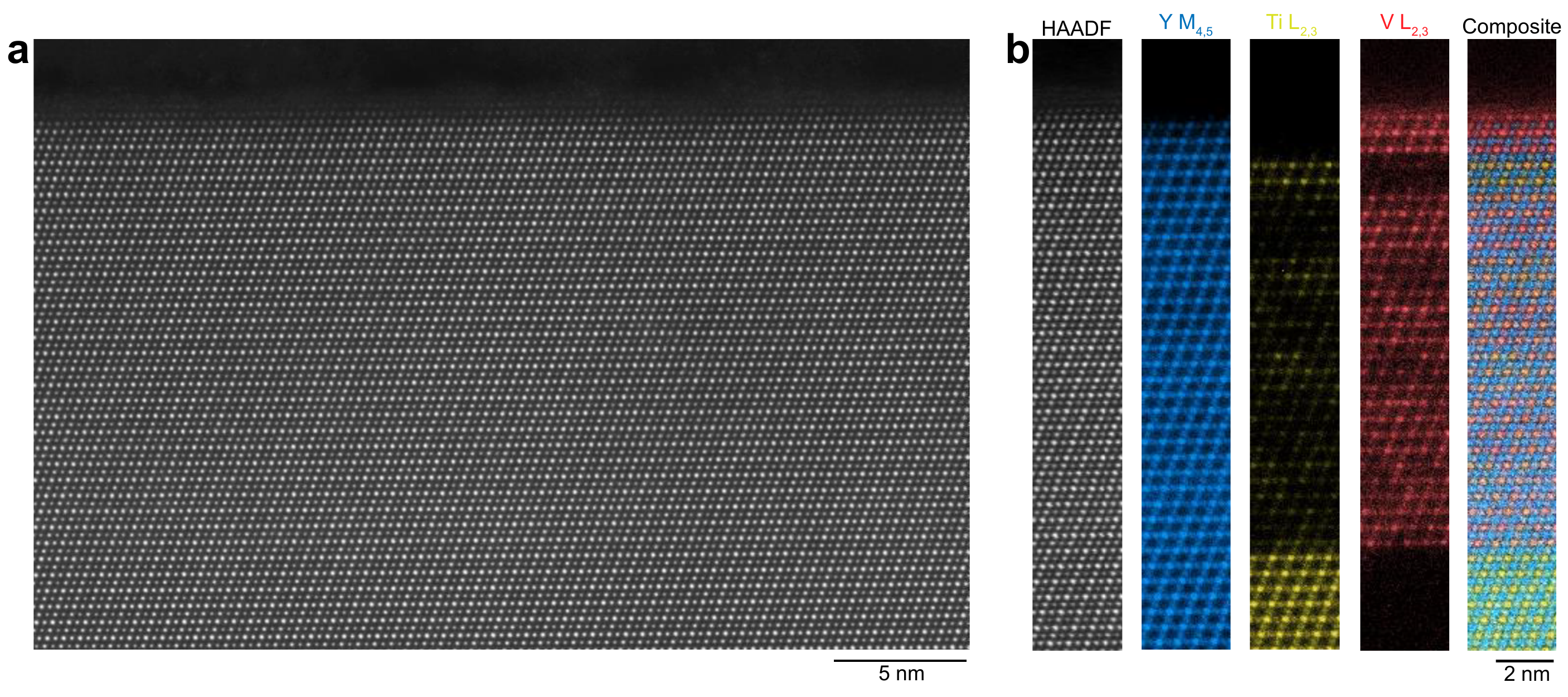}
\end{center}
    \captionof{figure}{STEM-EELS imaging of (Y$_2$Ti$_2$O$_7$)$_6$/(Y$_2$V$_2$O$_7$)$_6$ superlattice on Y$_2$Ti$_2$O$_7$. \textbf{a} A HAADF-STEM micrograph taken along $\langle1\overline10\rangle$ with collection angle 64-200 mrad (same conditions as Figure \ref{suppSL-STEM}a) \textbf{b} From left to right: a HAADF-STEM micrograph of the region of interest at approximately the same scale as (a), an EELS concentration map of yttrium, a map of titanium concentration, a vanadium concentration map, and a composite view showing the layering within the superlattice sample. Diffusion increases deeper within the film, disrupting the clear 6 monolayer repeat visible at the surface of the film.}
    \label{suppSL-EELS}

\begin{center}
    \includegraphics[width=\textwidth]{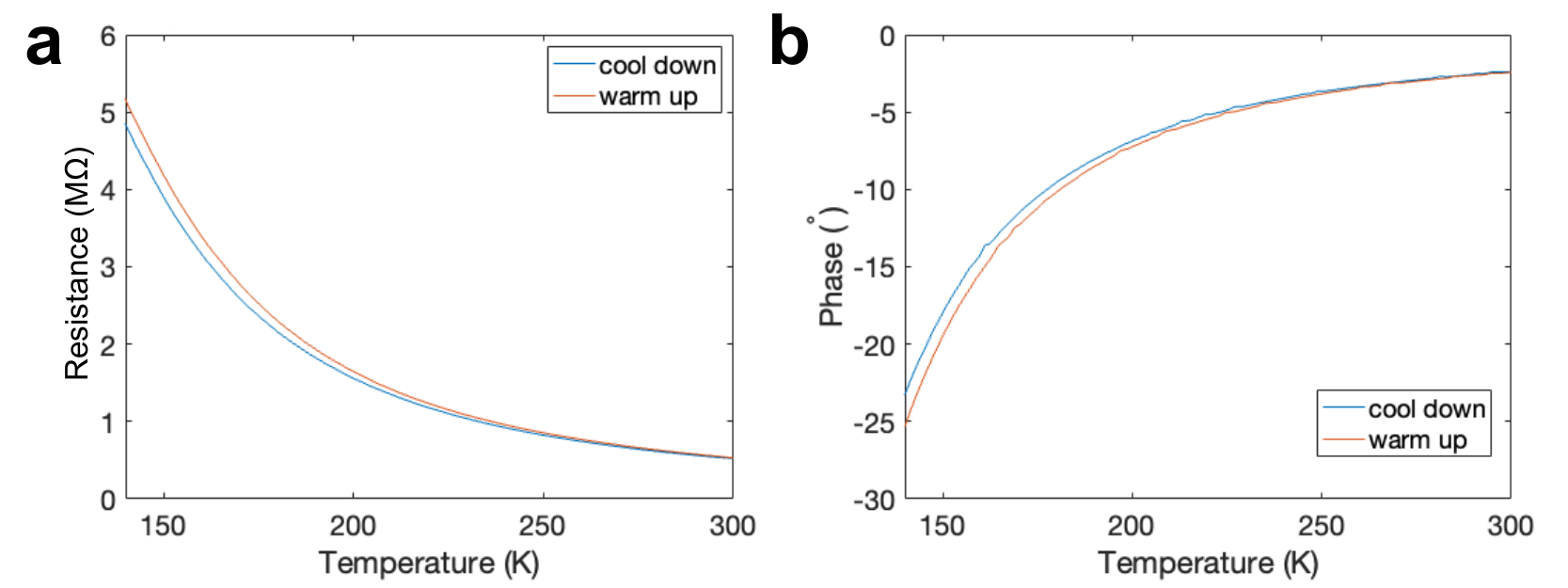}
\end{center}
    \captionof{figure}{Insulating transport behavior of a 60 atomic layer Y$_2$V$_2$O$_7$ film on Y$_2$Ti$_2$O$_7$. \textbf{a} Insulating behavior where resistance increases rapidly with decreasing temperature; an Arrenhius fit to the high temperature data estimates the band gap of about 50 meV, below bulk estimates of 0.2-0.33 eV; thinner films exhibit higher surface resistance at room temperature \textbf{b} The phase of the resistance measurement increases rapidly with decreasing temperature, which indicates the sample becomes highly resistive and the resistance measurement becomes less reliable}
    \label{suppRvsT}

\begin{center}
    \includegraphics[width=\textwidth]{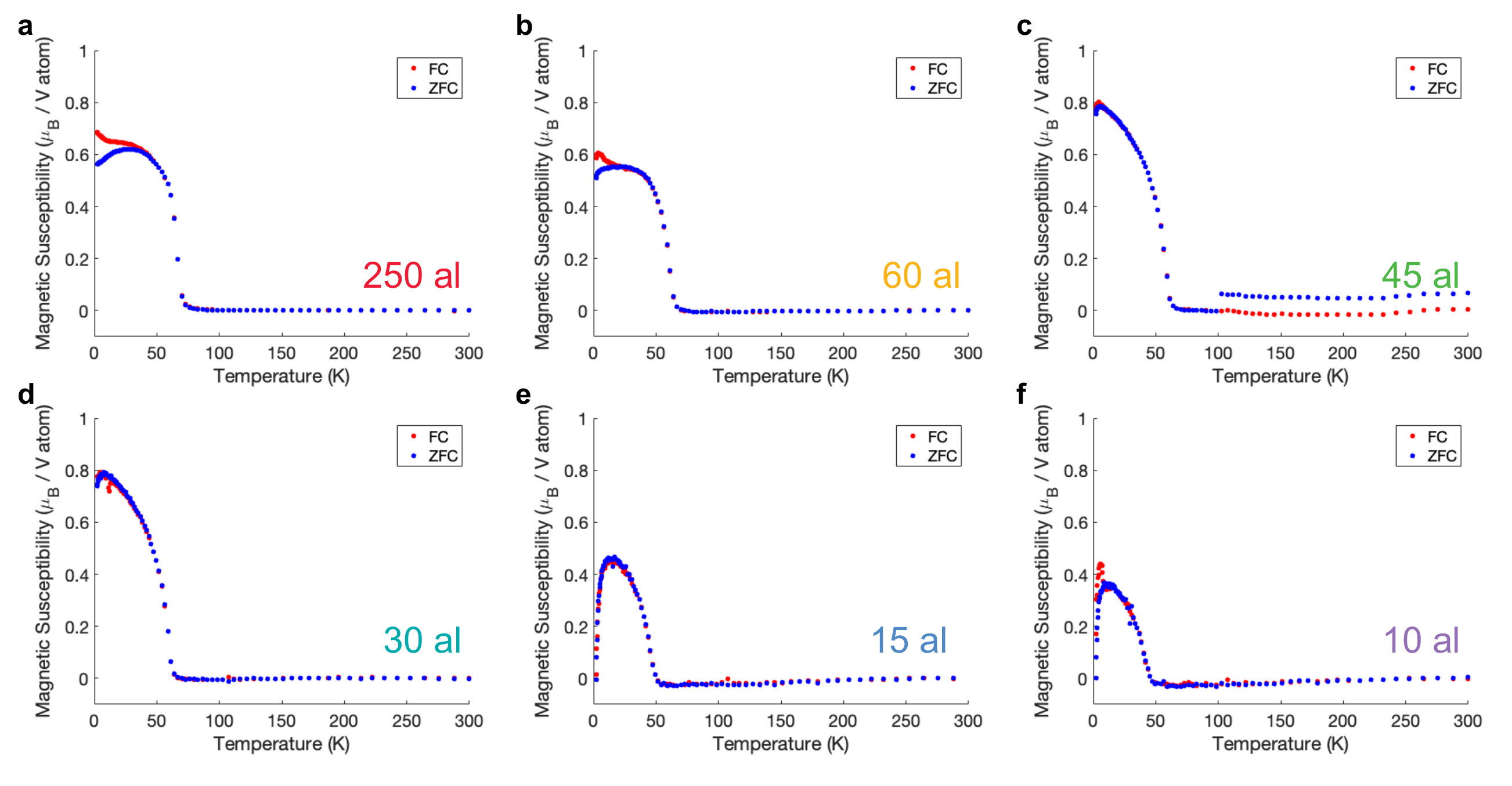}
\end{center}
    \captionof{figure}{Field-cooled versus zero-field-cooled (ZFC) magnetic susceptibility of Y$_2$V$_2$O$_7$ thickness series. \textbf{a-f} The zero-field-cooled (ZFC, blue) and 2 kOe field-cooled (FC, red) magnetic susceptibilities measured in a 500 Oe in-plane applied field for films with thickness (a) 250, (b) 60, (c) 45, (d) 30, (e) 15, and (f) 10 atomic layers of Y$_2$V$_2$O$_7$ on isostructural Y$_2$Ti$_2$O$_7$(111). The FC and ZFC curves diverge at low temperatures in the films at or above about 45 atomic layers thick, which suggests hysteresis opens in a changing applied field at these thicknesses and temperatures (as seen in Figure \ref{MHandHc})}
    \label{suppSuscFC-ZFC}

\begin{center}
    \includegraphics[width=0.5\textwidth]{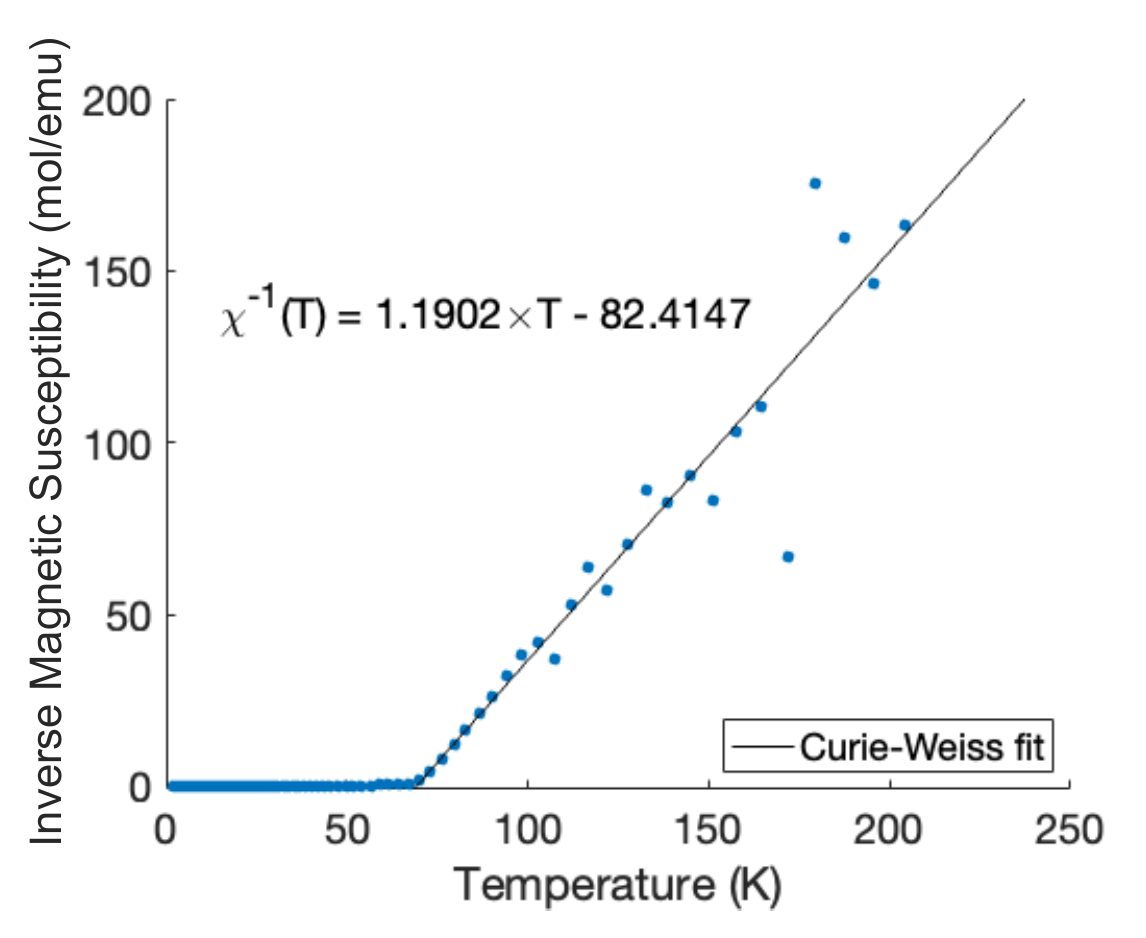}
\end{center}
    \captionof{figure}{Curie-Weiss fit of the inverse magnetic susceptibility of 250 atomic layer Y$_2$V$_2$O$_7$. A linear fit to the inverse field-cooled susceptibility between 80 and 200 K yields an estimated Curie-Weiss temperature of 69 K and $\mu_\text{eff}$ = 1.83 $\mu_\text{B}$/V atom.}
    \label{suppInvSusc}

\begin{center}
    \includegraphics[width=\textwidth]{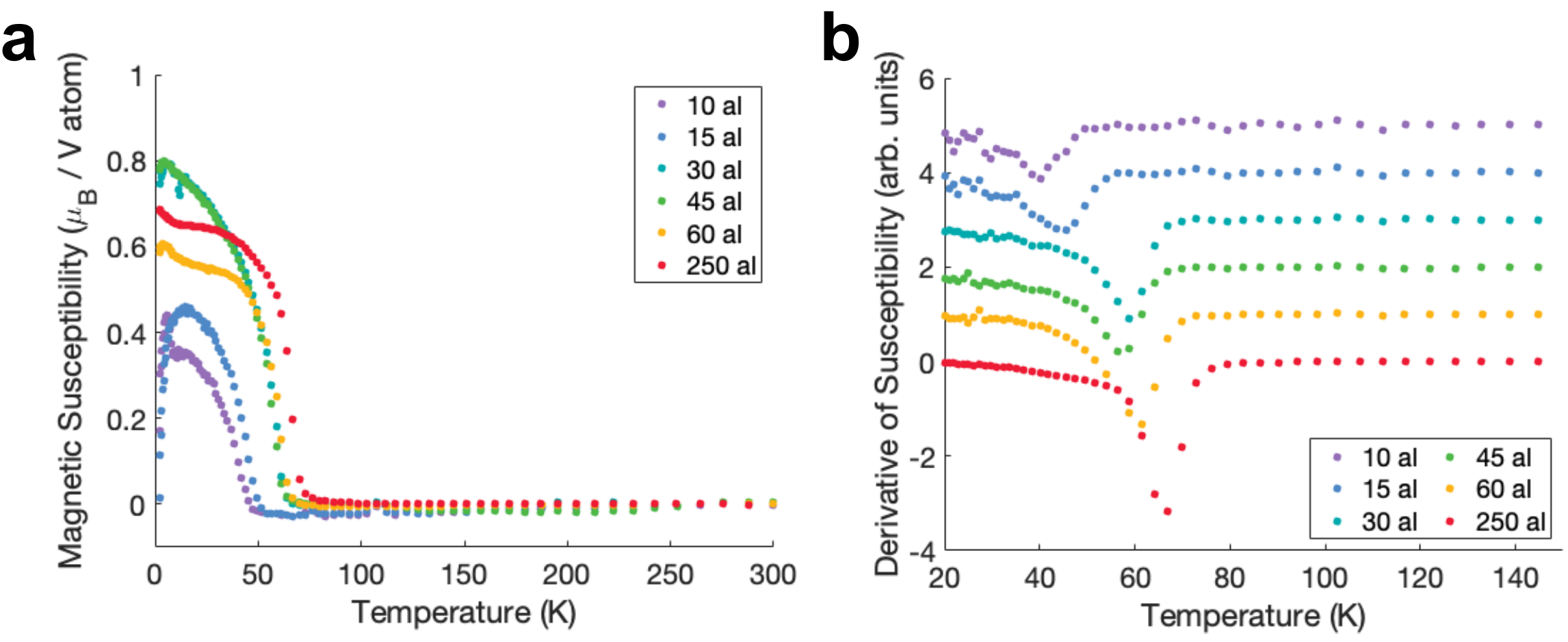}    
\end{center}
    \captionof{figure}{Volume normalization and derivative of the susceptibility of the Y$_2$V$_2$O$_7$ thickness series. \textbf{a} The substrate subtracted, diamagnetism corrected, volume normalized, field-cooled magnetic susceptibility of the Y$_2$V$_2$O$_7$ measured in a 500 Oe field shifted to align at high temperatures \textbf{b} The derivatives of the susceptibilities plotted in (a) with a vertical shift for clarity; T$_\text{c}$ is defined as the temperature where the absolute value of the derivative is maximized}
    \label{suppExtraSusc}

\begin{center}
    \includegraphics[width=0.5\textwidth]{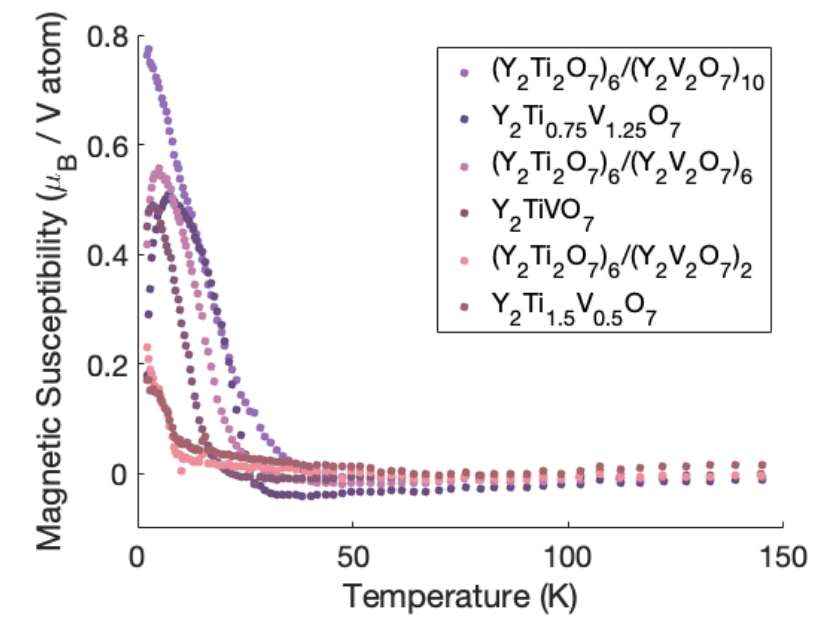}
\end{center}
    \captionof{figure}{Magnetic susceptibility of the (Y$_2$Ti$_2$O$_7$)$_6$/(Y$_2$V$_2$O$_7$)$_n$ superlattices. The substrate subtracted, diamagnetism corrected, volume normalized, field-cooled magnetic susceptibility of the (Y$_2$Ti$_2$O$_7$)$_6$/(Y$_2$V$_2$O$_7$)$_n$ and corresponding solid solution films on Y$_2$Ti$_2$O$_7$ measured in a 500 Oe field shifted to align at high temperatures}
    \label{suppSLSusc}

\begin{center}
    \includegraphics[width=\textwidth]{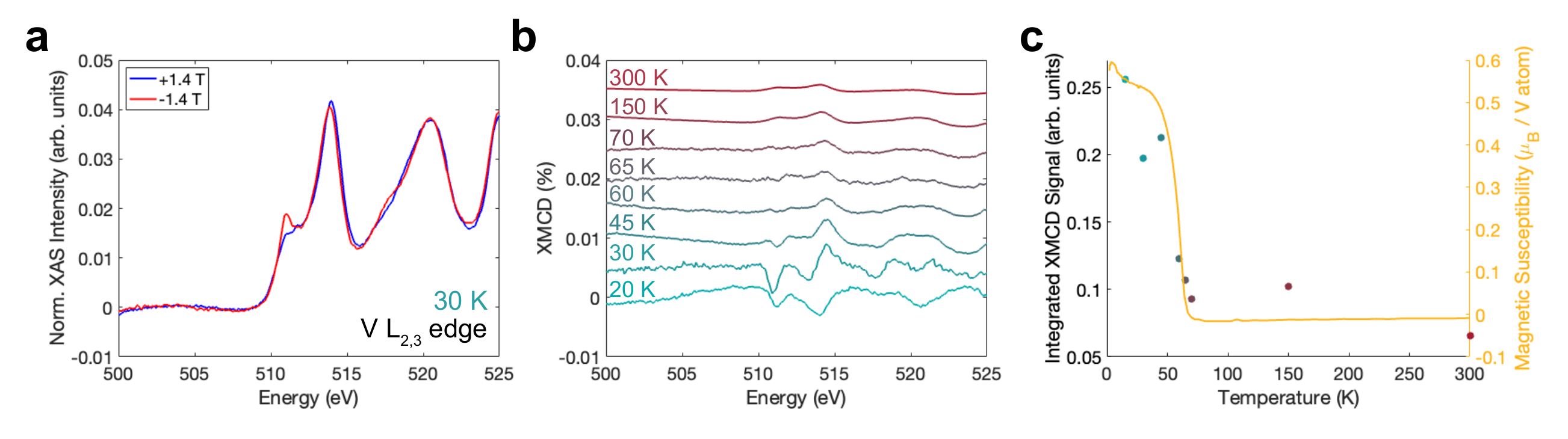}
\end{center}
    \captionof{figure}{X-ray circular magnetic dichroism (XMCD) of Y$_2$V$_2$O$_7$ on Y$_2$Ti$_2$O$_7$. \textbf{a} X-ray absorption spectroscopy of the V-$L_{2,3}$ edge of a 60 atomic layer Y$_2$V$_2$O$_7$ thin film under grazing incidence $\sigma^+$ polarized light at 30 K under $\pm 1.4$ T applied field; The XAS spectra show the expected shape for V$^{4+}$ in a nearly octahedral coordination environment \textbf{b} The XMCD signal at various temperatures extracted from the difference of XAS signals in $\pm 1.4$ T applied fields as in (a) \textbf{c} The correspondence between the integrated XMCD signal of the Y$_2$V$_2$O$_7$ film at different temperatures and the magnetic susceptibility measured with SQUID magnetometry.}
    \label{suppXMCD}

\begin{center}
    \includegraphics[width=0.8\textwidth]{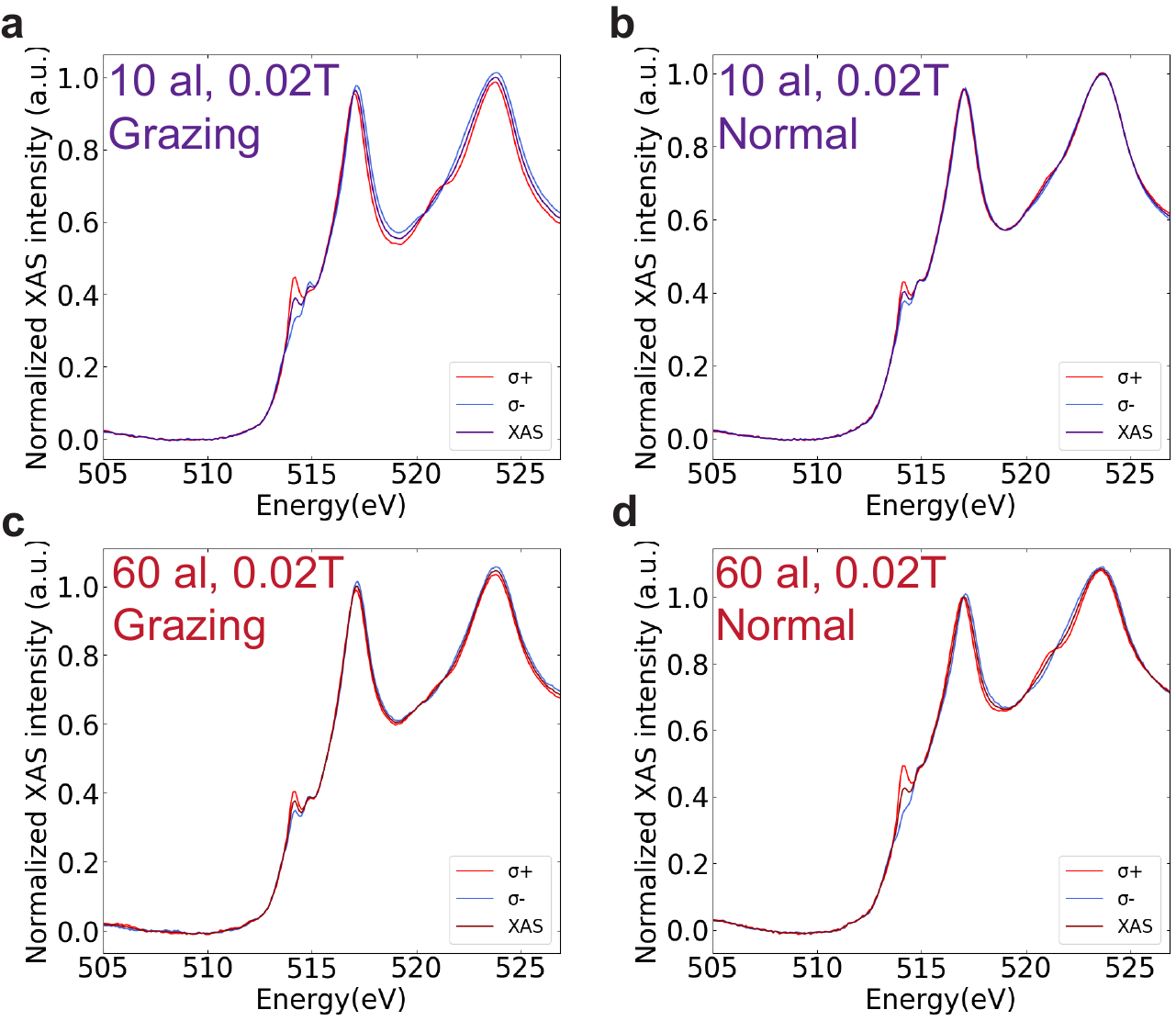}
\end{center}
    \captionof{figure}{X-ray absorption spectroscopy (XAS) of Y$_2$V$_2$O$_7$ on Y$_2$Ti$_2$O$_7$ measured at 20 K and 0.02 T \textbf{a, b} XAS of 10 atomic layer sample at grazing and normal incidences \textbf{c, d} XAS of 10 atomic layer sample at grazing and normal incidences}
    \label{suppAngleXAS}

\begin{center}
    \includegraphics[width=0.8\textwidth]{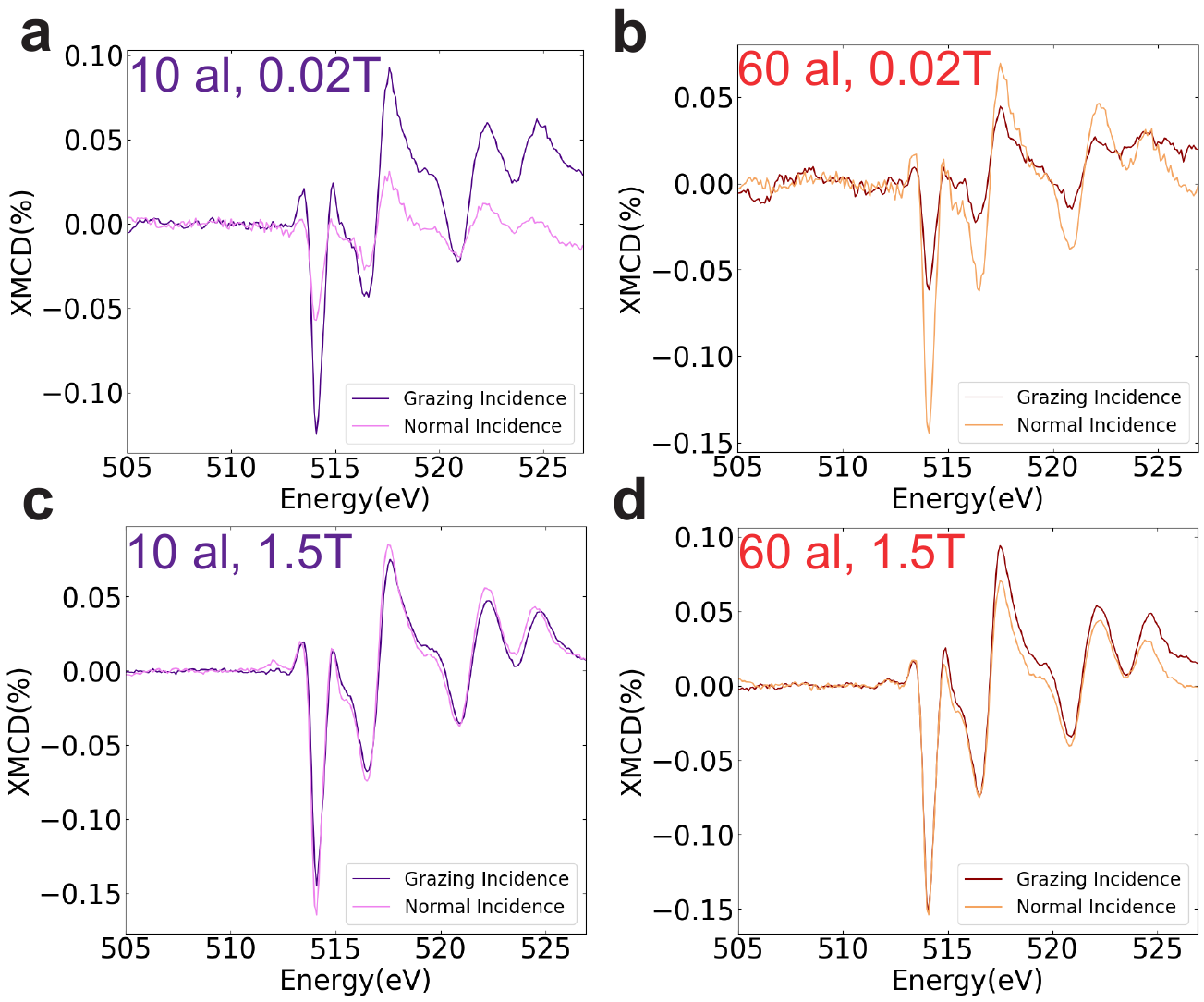}
\end{center}
    \captionof{figure}{Angle Dependent XMCD of Y$_2$V$_2$O$_7$ on Y$_2$Ti$_2$O$_7$ measured at 20 K. \textbf{a, b} XMCD signal at the V-$L_{2,3}$ edge of 10 atomic layer and 60 atomic layer films at 0.02 T measured at both normal (90$^\circ$) and grazing (20$^\circ$) incidences; a change in anisotropy is observed as stronger XMCD signal switches from grazing incidence in the 10 atomic layer film to normal incidence in the 60 atomic layer film \textbf{c, d} XMCD signal at the V-$L_{2,3}$ edge of 10 atomic layer and 60 atomic layer films at 1.5 T measured at both normal (90$^\circ$) and grazing (20$^\circ$) incidences; after saturation of the magnetic moments in the 1.5 T applied field, anisotropy is no longer clearly observed}
    \label{suppAngleXMCD}

\newpage
\begin{center}
    \includegraphics[width=0.8\textwidth]{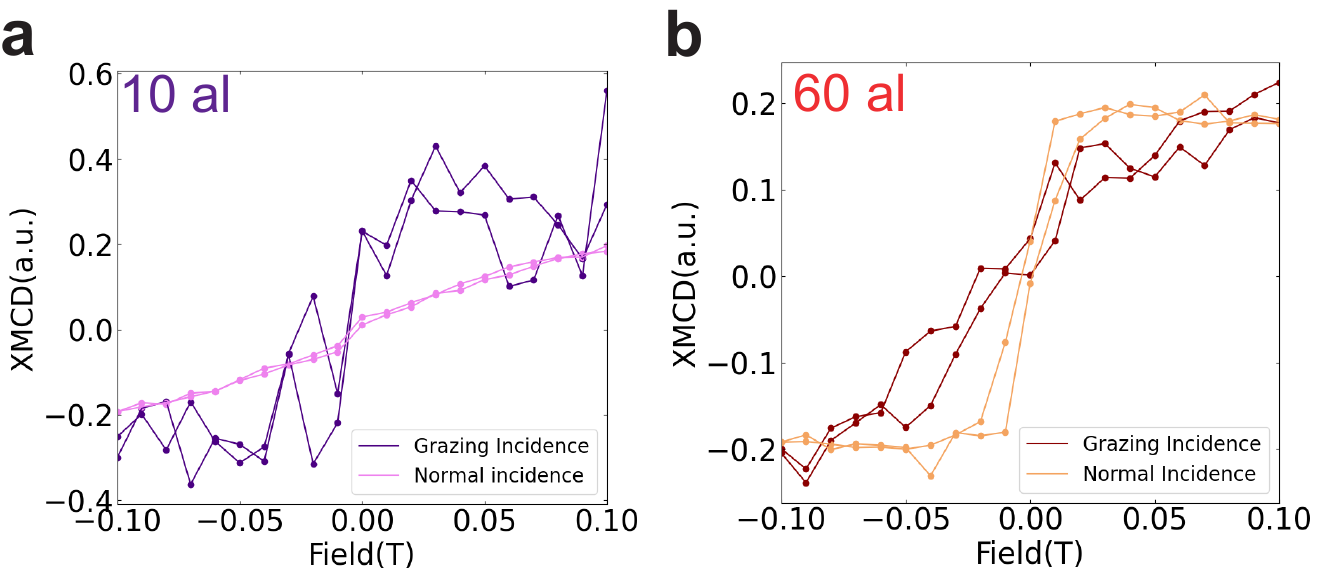}
\end{center}
    \captionof{figure}{Field dependent XMCD of Y$_2$V$_2$O$_7$ at 514.08 eV, where we observed the largest XMCD signal within the V-$L_{2,3}$ edge. Background scans were taken at 510 eV and subtracted from on-peak data \textbf{a} Field dependence from -0.1 to 0.1 T of 10 atomic layer film at grazing and normal incidence; we observe the grazing incidence reaches saturation at lower field \textbf{b} Field dependence from -0.1 to 0.1 T of 60 atomic layer film at grazing and normal incidence; we observe the normal incidence reaches saturation at lower field}
    \label{suppXMCDHystersis}

\begin{center}
    \includegraphics[width=\textwidth]{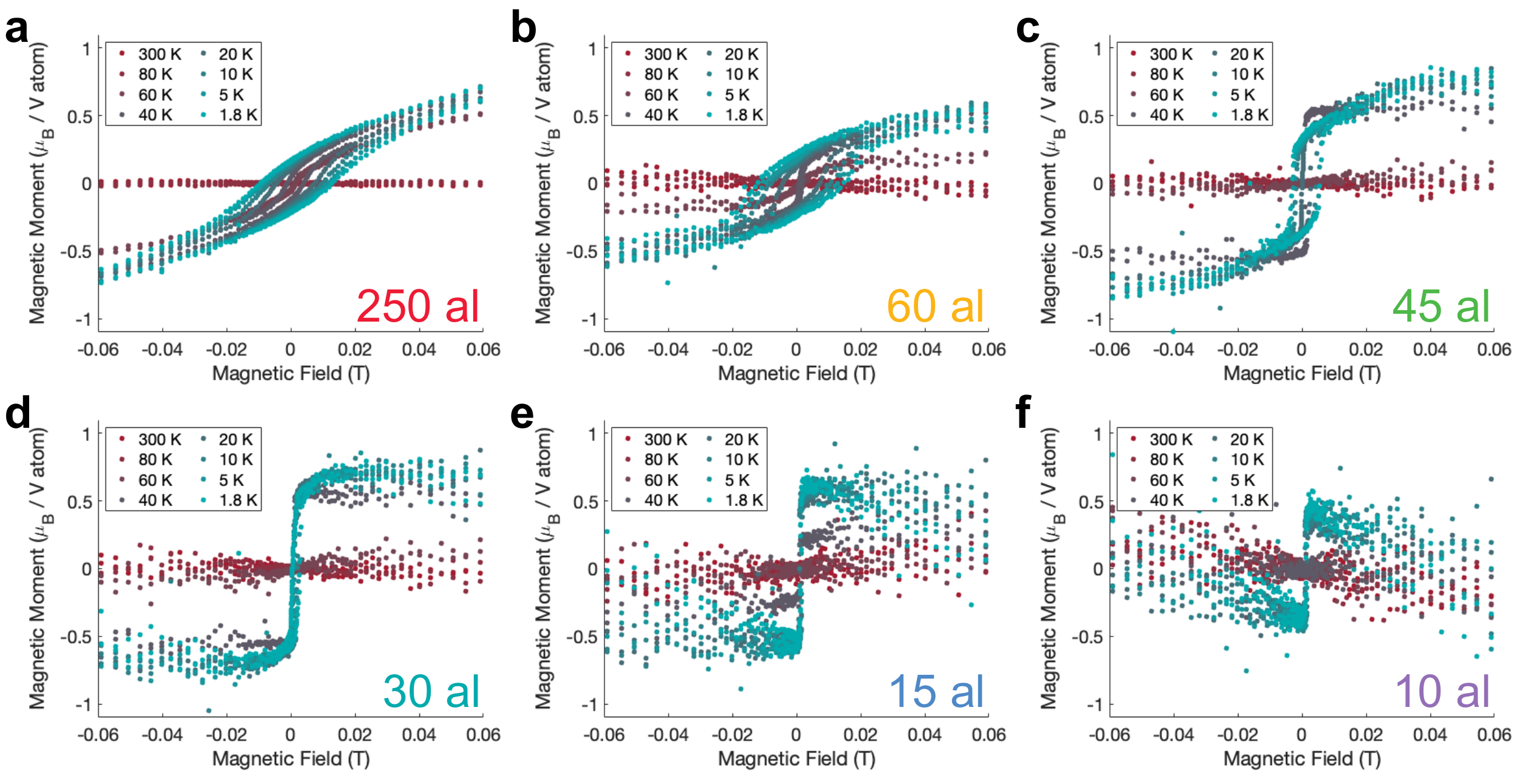}
\end{center}
    \captionof{figure}{Magnetization vs. in-plane applied field loops for Y$_2$V$_2$O$_7$ thickness series. \textbf{a-f} Temperature dependence of magnetization vs. applied field with field perpendicular to $\langle111\rangle$ (in-plane) for (a) 250, (b) 60, (c) 45, (d) 30, (e) 15, and (f) 10 atomic layers of Y$_2$V$_2$O$_7$ on Y$_2$Ti$_2$O$_7$; (b) is identical to Figure \ref{MHandHc}b in the main text}
    \label{suppSuppMH}

\begin{center}
    \includegraphics[width=\textwidth]{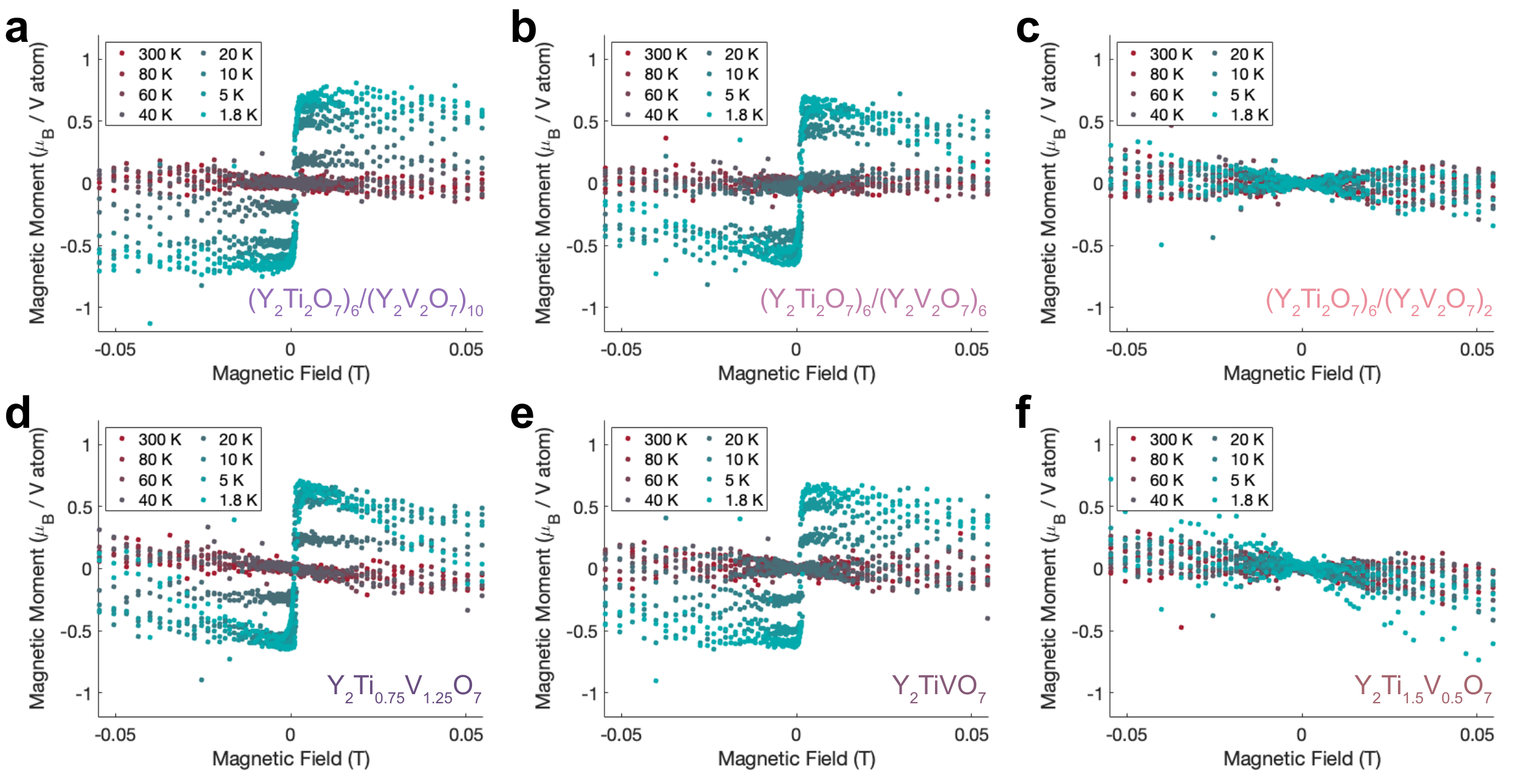}
\end{center}
    \captionof{figure}{Magnetization vs. in-plane applied field loops for (Y$_2$Ti$_2$O$_7$)$_6$/(Y$_2$V$_2$O$_7$)$_n$ superlattices. \textbf{a-f} Temperature dependence of magnetization vs. applied field with field perpendicular to $\langle111\rangle$ (in-plane) for (a) $n = 10$ and (d) corresponding cation disordered film Y$_2$Ti$_{0.75}$V$_{1.25}$O$_7$, (b) $n = 6$ and (e) corresponding Y$_2$TiVO$_7$, and (c) $n = 2$ and (f) corresponding Y$_2$Ti$_{1.5}$V$_{0.5}$O$_7$ on Y$_2$Ti$_2$O$_7$}
    \label{suppSLMH}

\begin{center}
    \includegraphics[width=\textwidth]{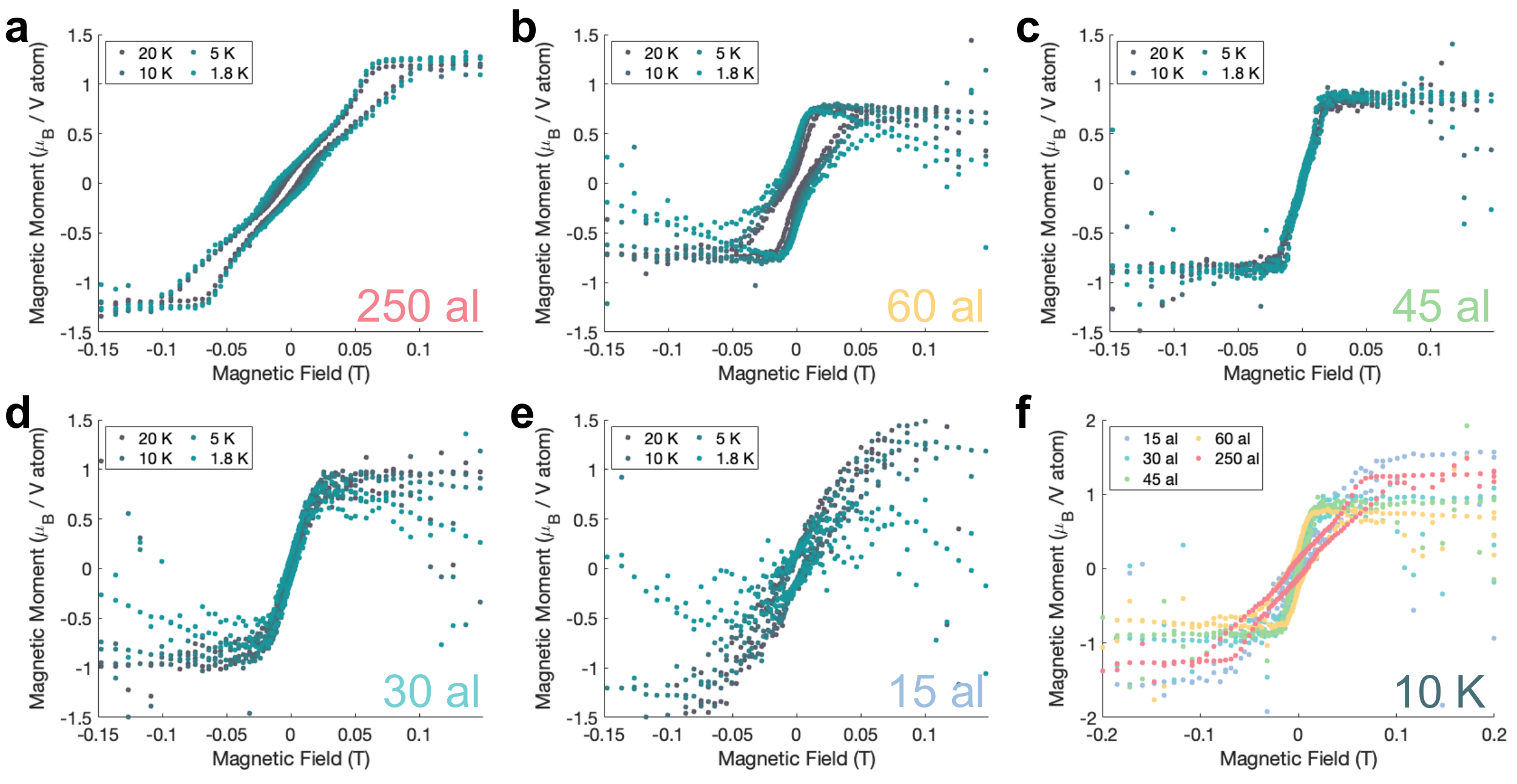}
\end{center}
    \captionof{figure}{Magnetization vs. out-of-plane applied field loops for Y$_2$V$_2$O$_7$ thickness series. \textbf{a-e} Temperature dependence of magnetization vs. applied field with field along $\langle111\rangle$ (out-of-plane) for (a) 250, (b) 60, (c) 45, (d) 30, and (e) 15 atomic layers of Y$_2$V$_2$O$_7$ on Y$_2$Ti$_2$O$_7$ \textbf{f} Thickness dependence of magnetization vs. applied field loops measured at 10 K with field along $\langle111\rangle$}
    \label{suppSuppMH-OOP}

\begin{center}
    \includegraphics[width= \textwidth]{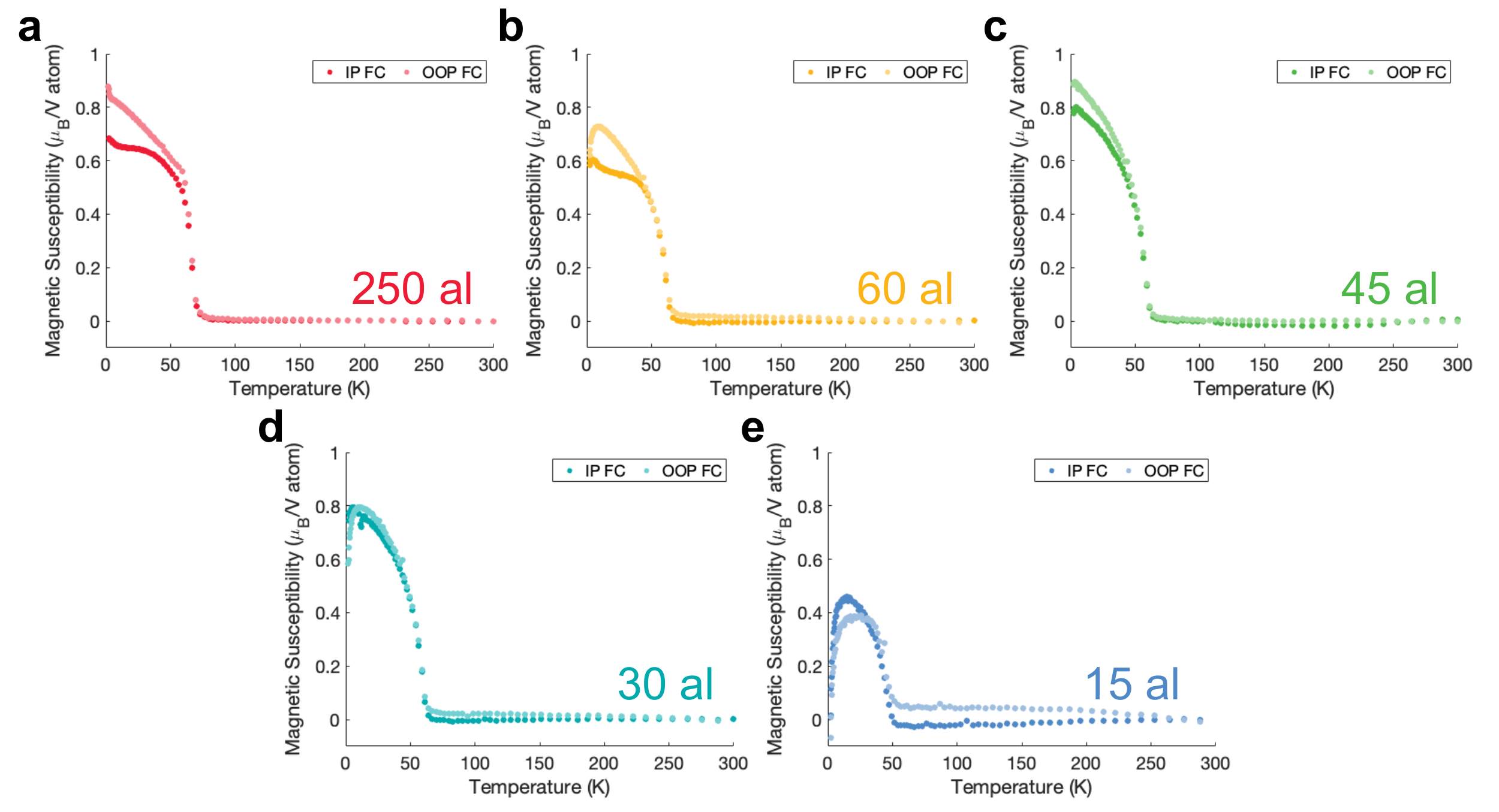}
\end{center}
    \captionof{figure}{Anisotropy of the magnetic susceptibility of Y$_2$V$_2$O$_7$ thickness series. \textbf{a-e} Magnetic susceptibility vs. temperature for a field applied within the plane of the film (IP, perpendicular to $\langle 111\rangle$) and out-of-plane (OOP, along $\langle 111\rangle$) for films with thickness (a) 250, (b) 60, (c) 45, (d) 30, and (e) 15 atomic layers, which show the same anisotropy trend as the magnetization vs. applied field loops in Figure \ref{aniso} }
    \label{suppSuscAniso}

\begin{center}
     \includegraphics[width=\textwidth]{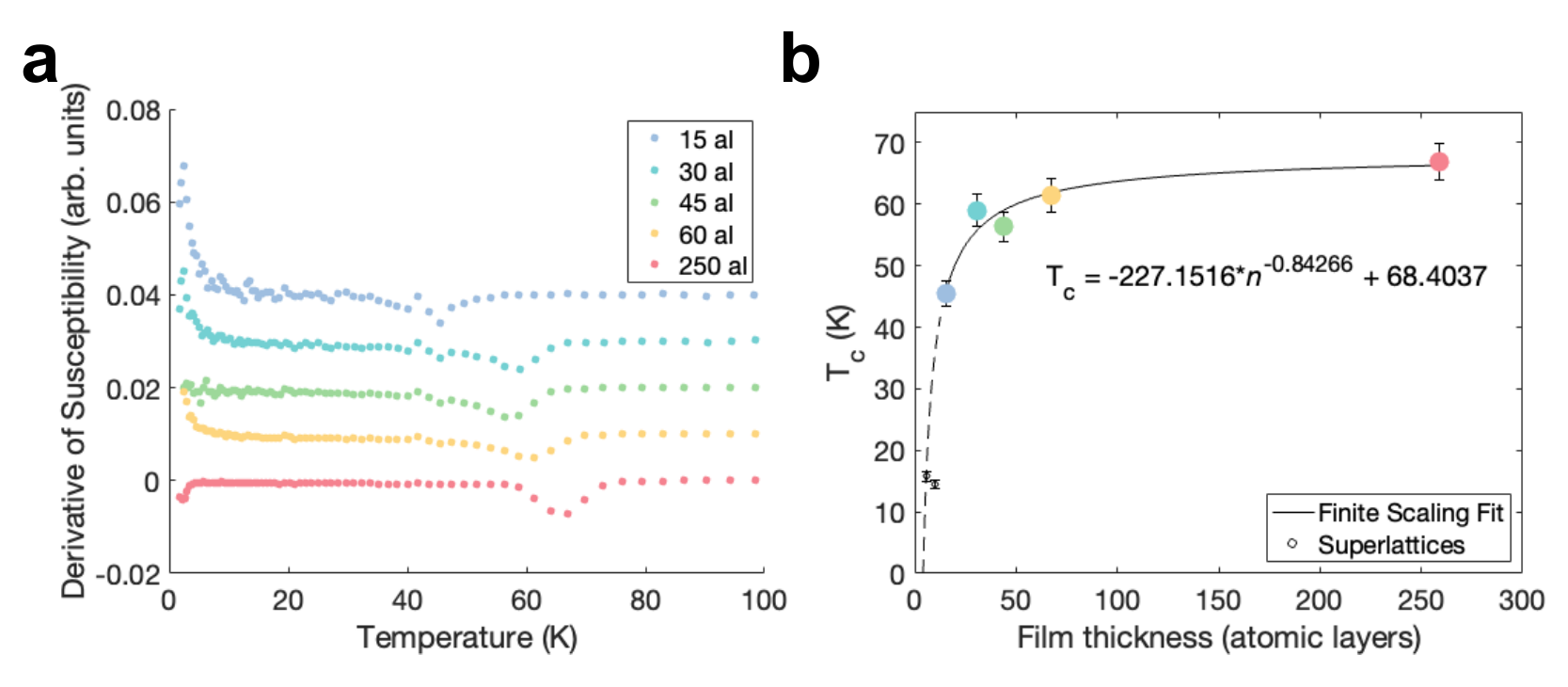}
\end{center}
    \captionof{figure}{Ferromagnetic transition temperatures of Y$_2$V$_2$O$_7$ thickness series in an out-of-plane field. \textbf{a} The derivatives of the magnetic susceptibilities measured in a magnetic field along $\langle111\rangle$ (plotted in lighter colors in Figure \ref{suppSuscAniso}) \textbf{b} The ferromagnetic transition temperatures vs. film thicknes ($n$, atomic layers) in an out-of-plane magnetic field defined by the temperature where the magnetic susceptibility has the largest slope magnitude; all transition temperatures and the fit to finite size are consistent with the results in an in-plane field (Figure \ref{SuscandTc}b)}
    \label{suppTcThickOOP}

\begin{center}
    \includegraphics[width=\textwidth]{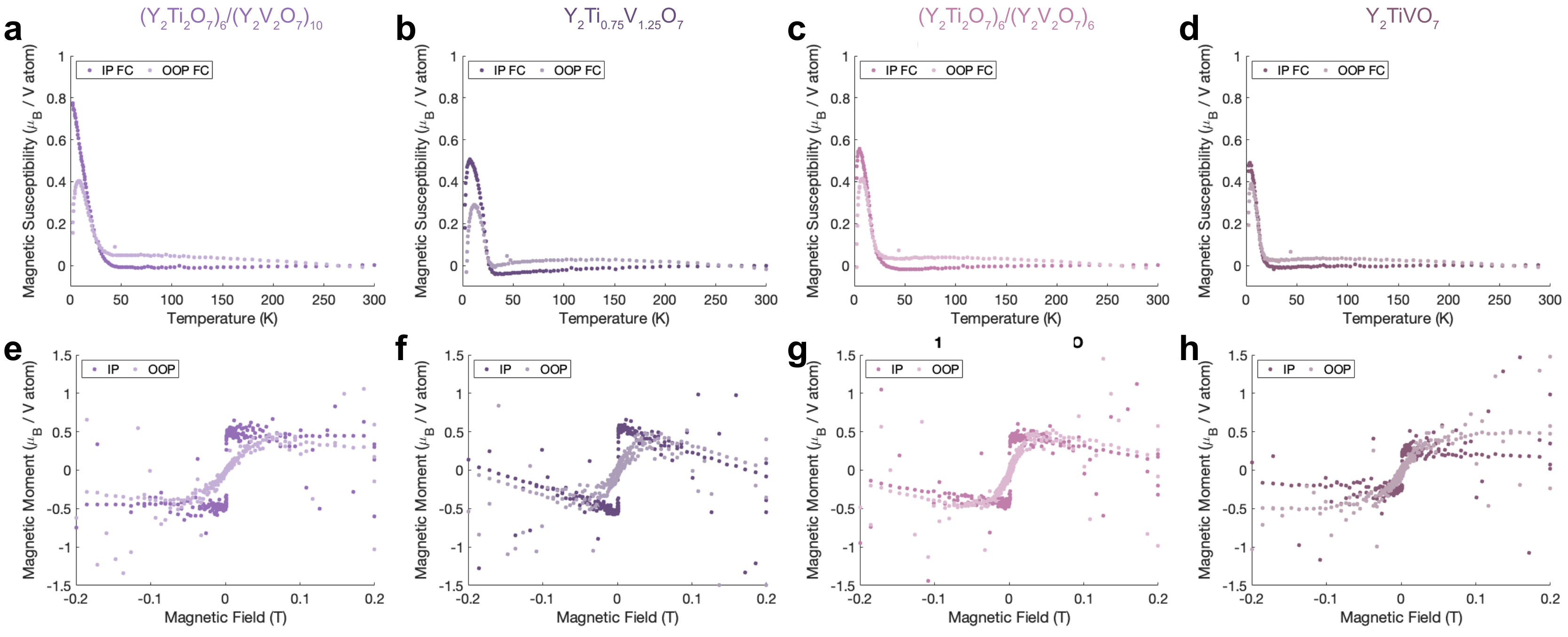}
\end{center}
    \captionof{figure}{Anisotropy of magnetism of (Y$_2$Ti$_2$O$_7$)$_6$/(Y$_2$V$_2$O$_7$)$_n$ superlattices. \textbf{a-d} Magnetic susceptiblity vs. temperature for a field applied within the plane of the film (IP, perpendicular to $\langle111\rangle$) and out-of-plane (OOP, along $\langle111\rangle$) for (a) $n = 10$, (b) the corresponding Y$_2$Ti$_{0.75}$V$_{1.25}$O$_7$, (c) $n = 6$, and (d) Y$_2$TiVO$_7$ \textbf{e-h} Magnetization vs. applied field loops at 10 K for a field applied in-plane and out-of-plane for the same films as the susceptibility above: (e) $n = 10$, (f) the corresponding Y$_2$Ti$_{0.75}$V$_{1.25}$O$_7$, (g) $n = 6$, and (h) Y$_2$TiVO$_7$ }
    \label{suppSLOOP}

\begin{center}
    \includegraphics[width=\textwidth]{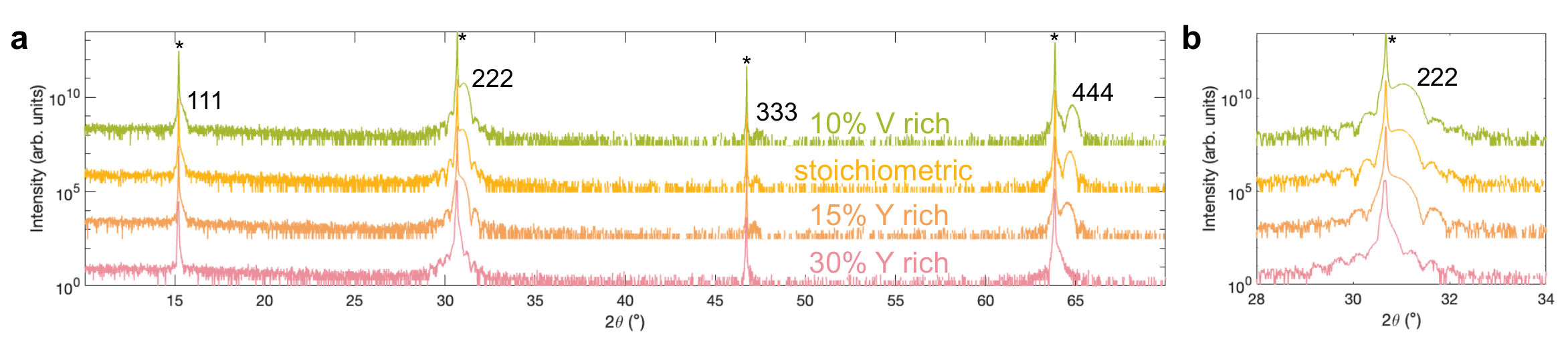}
\end{center}
    \captionof{figure}{X-ray diffraction of Y$_2$V$_2$O$_7$ on Y$_2$Ti$_2$O$_7$ with varying stoichiometry. \textbf{a} Full x-ray diffraction of a stoichiometry series of Y$_2$V$_2$O$_7$ showing a slight shifting of film peaks to higher angle with additional vanadium; asterisks indicate substrate diffraction peaks and lattice plane indices denote film peaks \textbf{b} Detail of 222 film peaks in (a)}
    \label{suppXRDstoich}

\begin{center}
   \includegraphics[width=\textwidth]{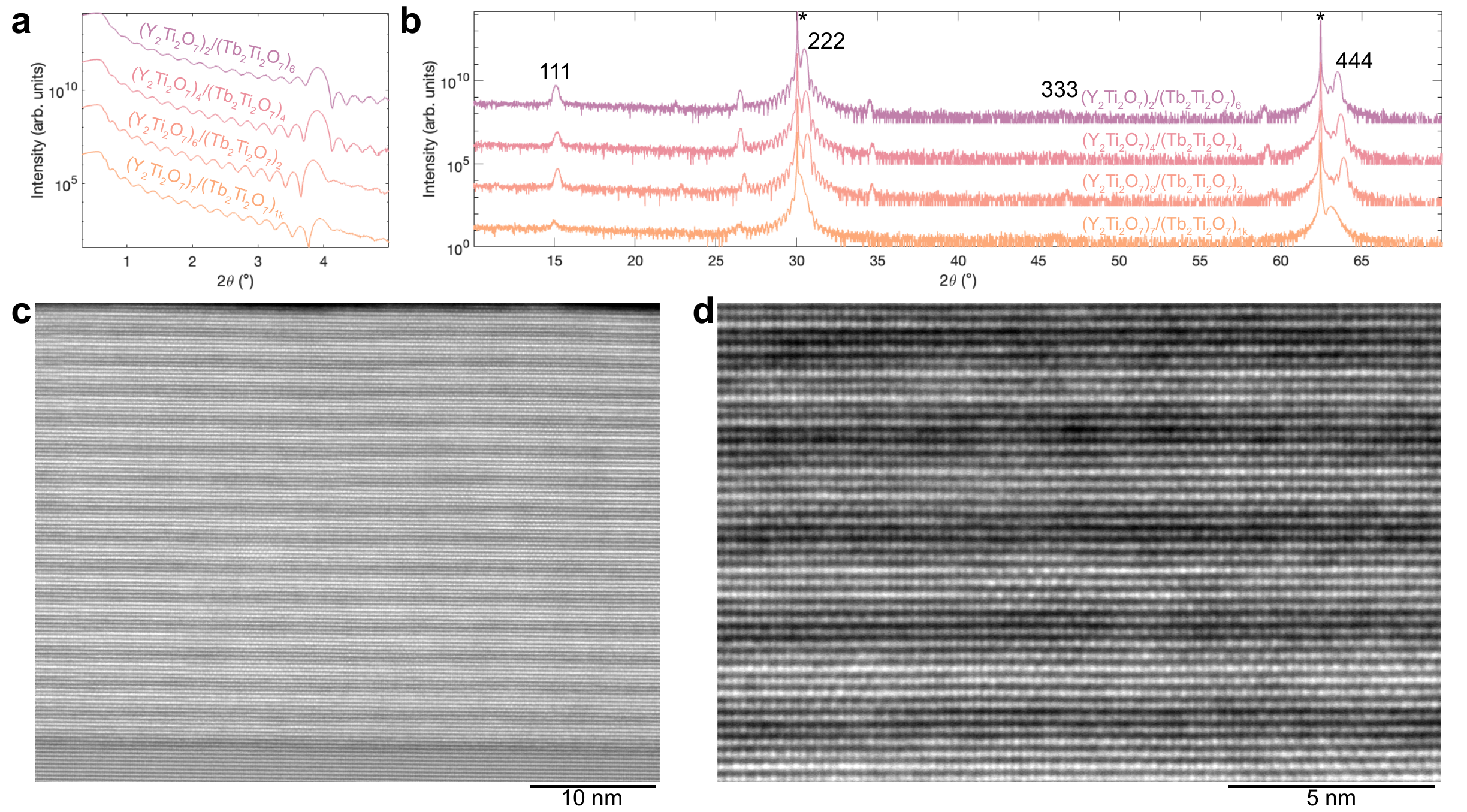} 
\end{center}
    \captionof{figure}{Characterization of (Y$_2$Ti$_2$O$_7$)$_m$/(Tb$_2$Ti$_2$O$_7$)$_n$) superlattices as a proof-of-concept. \textbf{a} X-ray reflectivity of superlattices indicating a smooth film with a clear low angle superlattice peak with a symmetry that depends on $m$ and $n$ and a constant position because $m+n=8$ for all films \textbf{b} X-ray diffraction showing high-quality pyrochlore films with clear superlattice satellite peaks about the typical pyrochlore peaks \textbf{c} A large field-of-view HAADF-STEM micrograph along $\langle11\overline{2}\rangle$ with clear contrast between the Tb$_2$Ti$_2$O$_7$ layers (brighter) and Y$_2$Ti$_2$O$_7$ layers (dimmer) \textbf{d} A higher magnification view of (c) showing the superlattice layering in more detail }
    \label{suppTTOsl}

\clearpage

\end{document}